\documentclass[12pt,prd,onecolumn,floatfix,nofootinbib]{revtex4}
\usepackage{amssymb}
\usepackage{amsmath}
\usepackage{graphicx,color,dcolumn,booktabs,bm}
\usepackage{longtable,lscape}
\usepackage{txfonts}
\usepackage{overpic}
\usepackage{indentfirst}
\usepackage{cases}
\usepackage{multirow}
\usepackage{epstopdf}
\usepackage{ulem}
\usepackage[colorlinks,
            citecolor=blue,
            anchorcolor=red,
            menucolor=red,
            linkcolor=red,
            filecolor=red,
            runcolor=red,
            urlcolor=blue,
            frenchlinks=false]{hyperref}

\allowdisplaybreaks
\begin{document}

\title{Masses and magnetic moments of hadrons with one and two open heavy
quarks: heavy baryons and tetraquarks}
\author{Wen-Xuan Zhang$^{1}$}
\email{zhangwx89@outlook.com}
\author{Hao Xu$^{1,2}$}
\email{xuh2020@nwnu.edu.cn}
\author{Duojie Jia$^{1,2}$ \thanks{}}
\email{jiadj@nwnu.edu.cn; corresponding author}
\affiliation{$^1$Institute of Theoretical Physics, College of Physics and Electronic
Engineering, Northwest Normal University, Lanzhou 730070, China \\
$^2$Lanzhou Center for Theoretical Physics, Lanzhou University, Lanzhou,
730000,China \\
}

\begin{abstract}
In this work, we compute masses and magnetic moments of the heavy baryons
and tetraquarks with one and two open heavy flavors in a unified framework
of MIT bag model. Using the parameters of MIT bag model, we confirm that an
extra binding energy, which is supposed to exist between heavy quarks ($c$
and $b$) and between heavy and strange quarks in literatures, is
required to reconcile light hadrons with heavy hadrons. Numerical
calculations are made for all light mesons, heavy hadrons with one and two
open heavy flavors, predicting the masses of doubly charmed baryons to be $%
M(\Xi _{cc})=3.604$ GeV, $M(\Xi _{cc}^{\ast })=3.714$ GeV, and that of the
strange isosinglet tetraquark $ud\bar{s}\bar{c}$ with $J^{P}=0^{+}$ to be $%
M\left( ud\bar{s}\bar{c},0^{+}\right) =2.934$ GeV. The state mixing due to
chromomagnetic interaction is shown to be sizable for the strange scalar
tetraquark $nn\bar{s}\bar{c}$.

{\normalsize PACS number(s):12.39Jh, 12.40.Yx, 12.40.Nn} {\normalsize Key
Words: Heavy baryons, heavy tetraquark, Mass, Magnetic moment}
\end{abstract}

\maketitle
\date{\today }


\section{Introduction}

Four years ago, the LHCb Collaboration at CERN discovered the first doubly
charmed baryon $\Xi _{cc}^{++}$ with $J^{P}={1/2}^{+}$ and measured its mass
to be $3621.40\pm 0.78$ MeV \cite{Aaij:2017ueg}. Later, the $\Xi _{cc}^{++}$
state was confirmed in the decay to $\Xi _{cc}^{+}\pi ^{+}$\cite{LhcXi:prl18}
and its lifetime, mass and production cross-section were subsequently
measured\cite{LhcXiLf:prl18,LhcXcc:cp20}. Containing two charmed quarks,
such a baryon provide a unique probe for quantum chromodynamics(QCD), the
gauge theory of strong interactions. In addition, the observation provides a
useful experimental information about strength of interaction between two
heavy quarks and enables us to further explore tetraquarks $QQ^{\prime }\bar{%
q}\bar{q}$ containing two open heavy quarks, which is allowed by QCD. See
Refs. \cite{Karliner:2017qjm,Eichten:2017ffp,Luo:2017eub} for instance.
Recently, LHCb Collaboration reported the first exotic state $X_{0}(2900)$
with open heavy flavors and mass $2866\pm 7$ MeV\cite{LHCb:2020kd}, which is
interpreted to be an isosinglet tetraquark $cs\bar{u}\bar{d}$ in Ref.~\cite%
{Karliner:2020vsi}. More recently, observation of a doubly charmed
tetraquark $T_{cc}^{++}$ is reported also by LHCb Collaboration\cite%
{TccPoly:2021}. These findings, among others, make it of interest to explore
doubly heavy(DH) hadrons in details. There exist extensive studies of DH
hadrons with various approaches, including potential quark model and bag
model \cite%
{Fleck:1989mb,Ebert:2004ck,Roberts:2007ni,Albertus:2006ya,Giannuzzi:2009gh,Bernotas:2008fv,Bernotas:2012nz,He:2004px,KR:2014gca,Liu:2018euh}%
, AdS/QCD approach \cite{Gutsche:2011vb,Gutsche:2017oro,Dosch:2016zdv,Nielsen:2018ytt,Dosch:2020hqm} and relativistic quark model \cite{Faessler:2006ft}.
See Refs \cite{Ali:2017jda,Liu:2019zoy} for recent reviews.

In identifying and/or finding these DH hadrons experimentally, it is helpful
to have a systematic estimate of masses and other properties of them within
an unified framework. For instance, a mass predictions \cite%
{Ebert:2004ck,KR:2014gca} of the doubly charmed baryon $\Xi _{cc}$, which
are larger about $100$ MeV than that measured in 2002 by the SELEX
Collaboration at Fermilab \cite{Moinester:2002uw}(awaiting confirmation),
helps LHCb Collaboration to search the $\Xi _{cc}^{++}$\cite{Aaij:2017ueg}
eventually.

In this work, we apply MIT bag model \cite{DeGrand:1975cf,Johnson:1975zp}
with chromomagnetic interaction and a strong coupling $\alpha _{s}$ running
with the bag radius to systematically study the open heavy baryons and
tetraquarks with one and two heavy flavors and compute the masses and other
static properties(magnetic moments, electric charge radii) of them. It is
confirmed that an extra binding energy between heavy quarks ($c$ and $b$) and
between heavy and strange quarks is required to reconcile light hadrons with
heavy hadrons. Computed results are compared to other calculations and in
consistent with the measured masses and other properties of light hadrons
and singly heavy baryons in their ground states(except for $\pi $). For the
$J^{P}={\frac{1}{2}}^{+}$ states of heavy baryons $\Xi_{c}$, $\Xi _{b}$,
$\Xi _{bc}$, $\Omega _{bc}$ and the heavy tetraquarks, the chromomagnetic
mixing is taken into account, and the respective mass splittings are
computed variationally.

It is well known that bag model \cite{DeGrand:1975cf} embodies two primary
features of quantum chromodynamics (QCD): asymptotic freedom at short
distance and confinement at long distance. The simple structure of the model
enables us to describe mesons($q\bar{q}$), baryons($qqq$) and even hadrons
made of multiquarks. In the past few decades, bag model has been applied to
describe the doubly heavy baryons\cite%
{Fleck:1989mb,He:2004px,Bernotas:2012nz} and multiquark hadrons, including
light exotic baryons with five and seven nonstrange quarks \cite%
{Strottman:1979qu}. In order to evaluate the masses of doubly heavy baryons,
a large running strong coupling $\alpha _{s}$ was applied in Ref. \cite%
{He:2004px}. A Coulomb-like interaction is derived between heavy quarks in a
bag in Refs.~\cite{Haxton:1980mc,Aerts:1980rf}.

This paper will be organized as follows. In Sec.~\ref{Method}, we review
some basic relations of MIT bag model, including chromomagnetic interaction
(CMI) among the quarks in bag. In Sec.~\ref{Parameters}, a systematic
numerical calculation is performed for the established light and singly
heavy(SH) baryons, with the optimal set of parameters obtained and the results
for masses and other properties reproduced. In Sec.~\ref{Hadrons}, we
present detailed predictions for masses and other properties for doubly
heavy baryons and the tetraquarks with one and two open heavy quarks. The
paper ends with summary and conclusions in Sec.~\ref{Discussions}.

\section{Method for MIT bag model with CMI}

\label{Method}

\subsection{Mass Formula}

Treating hadron as a spherical bag, MIT bag model provides an approach to
estimate masses and other properties of hadrons in their ground states\cite%
{Johnson:1975zp,DeGrand:1975cf}, in which the chromomagnetic interaction is
derived from the energy of a sphere-like gluon field interacting with quark
fields in bag \cite{DeGrand:1975cf}. The mass formula of hadron in MIT bag
model is,
\begin{equation}
M\left( R\right) =\sum_{i=n,s,c,b}n_{i}\omega _{i}+\frac{4}{3}\pi R^{3}B-%
\frac{Z_{0}}{R}+\langle \Delta H\rangle ,  \label{MBm}
\end{equation}%
\begin{equation}
\omega _{i}=\left( m_{i}^{2}+\frac{x_{i}^{2}}{R^{2}}\right) ^{1/2},
\label{freq}
\end{equation}%
where the first term is the kinematic energy of all quarks in bag with
radius $R$, the second is the volume energy of bag with bag constant $B$,
the third is the zero-point-energy (ZPE) with coefficient $Z_{0}$, and $%
\langle \Delta H\rangle $ is the short-range interaction among quarks in
bag, which we will address in this work. Here in Eq. (\ref{MBm}), $n_{i}$ is
number of quark or antiquark in bag with mass $m_{i}$ and flavor $i$, where $%
i$ can be the light nonstrange quarks $n=u,d$, the strange quark $s$, the
charm quark $c$ and the bottom quark $b$. The value of $R$ is to be
determined variationally, and the dimensionless parameters $x_{i}=x_{i}(mR)$
are related to the bag radius $R$ by an transcendental eigen-equation
\begin{equation}
\tan x_{i}=\frac{x_{i}}{1-m_{i}R-\left( m_{i}^{2}R^{2}+x_{i}^{2}\right)
^{1/2}}.  \label{transc}
\end{equation}

The interaction energy $\langle \Delta H \rangle=B_{EB}+ M_{CMI}$ is
composed of two energy terms:

(1) The spin-independent binding energy $B_{EB}$, due mainly to the
short-range chromoelectric interaction between quarks (and/or antiquarks).
Owing to its smallness for the relativistic light quarks $n$($=u,d$), this
energy, scales mainly as $-\sum \alpha _{s}/r_{ij}$, becomes sizable when
both of two quarks $i$ and $j$ are massive and moving nonrelativistically.
In present work, we treat this energy as sum of the pair binding energies $%
B_{QQ^{\prime }}$ ($B_{Qs}$) between heavy quarks and between heavy quark $Q$
and strange quark $s$ \cite{KR:2014gca,Karliner:2017elp,Karliner:2018bms}.
The net effect for this chromoelectric interaction amounts to introduction
of five binding energies $B_{cs}$, $B_{cc}$, $B_{bs}$, $B_{bb}$ and $B_{bc}$
for any quark pair in color configuration $\boldsymbol{\bar{3}}_{c}$, which
are extractable from heavy mesons and can be scaled to other color configurations.

(2) The chromomagnetic interaction energy, due to perturbative gluon
exchange between quarks (antiquarks) $i$ and $j$,
\begin{equation}
M_{CMI}=-\sum_{i<j}\left( \boldsymbol{\lambda _{i}}\cdot \boldsymbol{\lambda
_{j}}\right) \left( \boldsymbol{\sigma _{i}}\cdot \boldsymbol{\sigma _{j}}%
\right) C_{ij},  \label{CMI}
\end{equation}%
with $\boldsymbol{\lambda }_{i}$ the Gell-Mann matrices, $\boldsymbol{\sigma
}_{i}$ the Pauli matrices, and $C_{ij}$ the CMI parameter. In MIT bag model,
the parameters $C_{ij}$ are given by
\begin{equation}
C_{ij}=3\frac{\alpha _{s}\left( R\right) }{R^{3}}\bar{\mu}_{i}\bar{\mu}%
_{j}I_{ij},  \label{Cij}
\end{equation}%
\begin{equation}
\bar{\mu}_{i}=\frac{R}{6}\frac{4\alpha _{i}+2\lambda _{i}-3}{2\alpha
_{i}\left( \alpha _{i}-1\right) +\lambda _{i}},  \label{muBari}
\end{equation}%
\begin{equation}
I_{ij}=1+2\int_{0}^{R}\frac{dr}{r^{4}}\bar{\mu}_{i}\bar{\mu}%
_{j}=1+F(x_{i},x_{j}),  \label{Iij}
\end{equation}%
where $\alpha _{i}\equiv \omega _{i}R$, $\lambda _{i}\equiv m_{i}R$, $\alpha
_{s}\left( R\right) $ is the running strong coupling, $\bar{\mu}_{i}$ is the
reduced magnetic moment without electric charge, and $I_{ij}$ and $%
F(x_{i},x_{j})$ are rational functions of $x_{i}$ and $x_{j}$, given
explicitly by\cite{DeGrand:1975cf}.
\begin{equation}
\begin{aligned} F(x_{i}, x_{j})={\left(x_{i} {\rm
sin}^{2}x_{i}-\frac{3}{2}y_{i}\right)}^{-1} {\left(x_{j} {\rm
sin^{2}}x_{j}-\frac{3}{2}y_{j}\right)}^{-1} \\
\left\{-\frac{3}{2}y_{i}y_{j}-2x_{i}x_{j} {\rm sin}^{2}x_{i} {\rm
sin}^{2}x_{j} +\frac{1}{2}x_{i}x_{j}\left[2x_{i} {\rm Si}(2x_{i})
\right.\right.\\ \left.\left. +2x_{j} {\rm Si}(2x_{j}) -(x_{i}+x_{j}) {\rm
Si}(2(x_{i}+x_{j})) \right.\right.\\ \left.\left. -(x_{i}-x_{j}) {\rm
Si}(2(x_{i}-x_{j})) \right] \right\}, \end{aligned}  \label{F}
\end{equation}%
where $y_{i}=x_{i}-\mathrm{cos}(x_{i})\mathrm{sin}(x_{i})$, $x_{i}$ is the root
of Eq. (\ref{transc}) for a given $m_{i}R$, and
\begin{equation}
\mathrm{Si}(x)=\int_{0}^{x}\ \frac{{\rm sin}(t)}{t}{\rm d}t.
\end{equation}

\vspace{-1.1cm}
\setlength{\abovecaptionskip}{-1.0cm}

\begin{figure}[!ht]
\begin{center}
\includegraphics[width=0.77\textwidth]{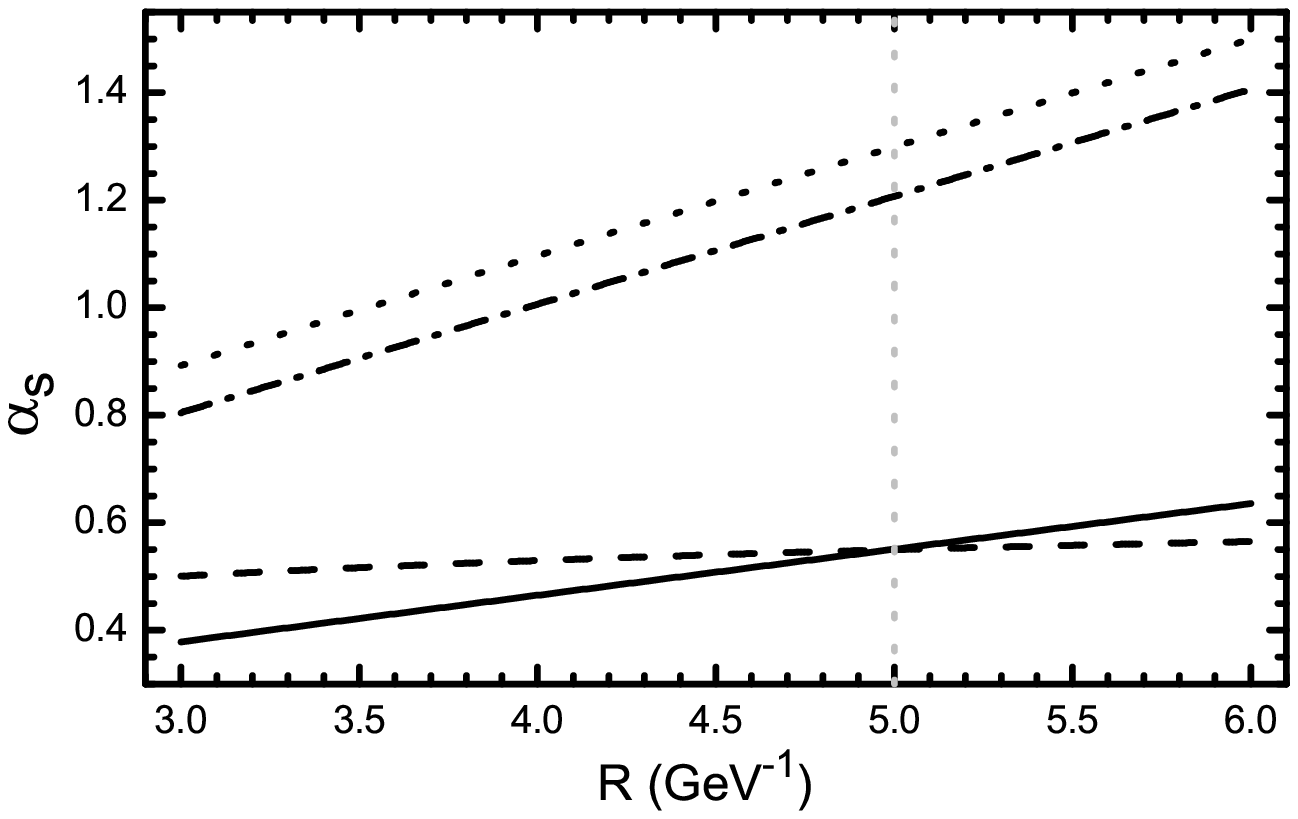}
\end{center}
\caption{Four running behaviors of strong coupling. Bag radius $R$ ranges
from $3\,\mathrm{GeV}^{-1}$ to $6\,\mathrm{GeV}^{-1}$, and the standard
radius is set to be $5\,\mathrm{GeV}^{-1}$ ($\approx 1\,\mathrm{fm}$) for
checking. The solid line represents our result (\protect\ref{alphaS-mine})
while the dashed line corresponds to Eq.~(\protect\ref{alphaS}) with $%
\protect\gamma=2.847$, $\Lambda_{QCD}=0.281\,\mathrm{GeV}$ and $w=2\protect%
\pi n/9$. The dotted line shows Eq.~(\protect\ref{alphaS}) with $\protect%
\gamma=1$, $\Lambda_{QCD}=0.281\,\mathrm{GeV}$ and $w=2\protect\pi n/9$. The
dotdashed line indicates that of Ref.~\protect\cite{He:2004px}. All four
behaviors adopt $n=1$.}
\label{fig:alphaS}
\end{figure}

In some applications of bag model \cite%
{Haxton:1980mc,Aerts:1980rf,Carlson:1982er,He:2004px,Fleck:1989mb}, the
parameter $\alpha _{s}$ takes a logarithmic form
\begin{equation}
\alpha _{s}(R)=\frac{w}{\mathrm{ln}\left[ \gamma +{\left( R\Lambda
_{QCD}\right) }^{-n}\right] },  \label{alphaS}
\end{equation}%
where $w$ and $n$($=1$ or $2$) are the parameters, $\Lambda _{QCD}$ is the
QCD scale($0.2\sim 0.5$ GeV), and $\gamma $ is prefactor used to avoid infrared
divergence. Similar to Ref.~\cite{He:2004px}, we take $w=0.296$, $\Lambda
_{QCD}=0.281$ GeV, $\gamma =1$ and $n=1$ to set
\begin{equation}
\alpha _{s}(R)=\frac{0.296}{\mathrm{ln}\left[ 1+{\left( 0.281R\right) }^{-1}%
\right] },  \label{alphaS-mine}
\end{equation}%
which is plotted in Fig.~\ref{fig:alphaS}. Among four lines showing the
running of $\alpha _{s}$ in the plot, the solid line shows notable variation
and corresponds to a relative lower value of $\alpha _{s}$.

\renewcommand{\tabcolsep}{0.39cm} \renewcommand{\arraystretch}{1.2}
\begin{table}[!htb]
\setlength{\abovecaptionskip}{0.4cm}

\caption{Bag radius $R$ (in GeV$^{-1}$) and mass prediction $M_{bag}$ (in
GeV) obtained from this work ($Z_{0}=1.83$) and original MIT bag model for
light hadrons, compared to the measured mass $M_{exp}$ (in GeV) being
isospin-averaged.}
\label{tab:lighthadron}%
\begin{tabular}{cccccc}
\hline\hline
\multirow{2}{*}{State} & \multicolumn{2}{c}{MIT bag\cite{Johnson:1975zp}} &
\multicolumn{2}{c}{\textrm{This\ work}} & \multirow{2}{*}{$M_{exp}$
\cite{Tanabashi:D18}} \\ \cline{2-3}\cline{4-5}
& $R$ & $M_{bag}$ & $R$ & $M_{bag}$ &  \\ \hline
$N$ & 5.00 & 0.938 & 5.22 & 0.932 & 0.939 \\
$\Delta$ & 5.48 & 1.233 & 5.33 & 1.241 & 1.232 \\
$\Lambda$ & 4.95 & 1.105 & 5.26 & 1.096 & 1.116 \\
$\Sigma$ & 4.95 & 1.144 & 5.22 & 1.137 & 1.193 \\
$\Xi$ & 4.91 & 1.289 & 5.27 & 1.282 & 1.318 \\
$\Sigma^{\ast}$ & 5.43 & 1.382 & 5.38 & 1.383 & 1.385 \\
$\Xi^{\ast}$ & 5.39 & 1.529 & 5.42 & 1.529 & 1.533 \\
$\Omega$ & 5.35 & 1.672 & 5.46 & 1.677 & 1.672 \\
$\pi$ & 3.34 & 0.280 & 4.31 & 0.348 & 0.137 \\
$\omega$ & 4.71 & 0.783 & 4.55 & 0.776 & 0.783 \\
$K$ & 3.26 & 0.497 & 4.34 & 0.561 & 0.496 \\
$K^{\ast}$ & 4.65 & 0.928 & 4.63 & 0.918 & 0.894 \\
$\phi$ & 4.61 & 1.068 & 4.70 & 1.064 & 1.019 \\ \hline\hline
\end{tabular}%
\end{table}

Given the parameter values of the quark mass $m_{i}$, bag constant $B$, the
ZPE coefficient $Z_{0}$ and strong coupling constant $\alpha _{s}(R)$
depending on bag radius $R$, one can apply variational method to determine
the respective bag radius $R$ for each hadron and the respective $x_{i}$
through Eq. (\ref{transc}). Then, it is straightforward to use Eqs. (\ref%
{MBm}),(\ref{freq}) and (\ref{CMI}) to compute the ground-state masses and
other static properties (magnetic moments, the charge radius) of the hadrons
ranging from the light hadrons to heavy tetraquarks. The computed results
for the light hadrons are listed in Table \ref{tab:lighthadron}, compared to
that predicted by original MIT bag model.

We stress that for a given hadronic state there is in principle a unique set
of the solution $x_{i}$ and $R$ corresponding to respective bag dynamics, as
indicated by our computation. Owing to $(x_{i},R)$-dependence of the $%
\langle \Delta H \rangle$, a simple and analytic mass formula is lacking for
the hadrons with chromomagnetic-mixing since for that purpose one has to
first diagonalize the CMI matrices before the variational analysis, which
amounts to a higher-order algebraic equations.

\subsection{Chromomagnetic Interaction}

In evaluating the spin-dependent mass due to the CMI (\ref{CMI}), in which $%
\boldsymbol{\lambda _{i}}$ should be replaced by $-\boldsymbol{\lambda
_{i}^{\ast }}$ for an antiquark, one has to diagonalize the CMI matrix for
given hadron multiplets with certain spin-parity $J^{P}$ to give the
respective mass splittings \cite{Luo:2017eub} within the multiplets. For
this, we list all the flavor-spin-color wavefunctions of hadrons including
tetraquarks considered in this work, and present relevant formulas of the
color and spin factors for them.

\textbf{Mesons}: The color wavefunction $\phi ^{M}=\left\vert q_{1}\bar{q}%
_{2}\right\rangle $ can be one of two spin states (of vector and scalar
like):
\begin{equation}
\chi _{1}^{M}={\left\vert q_{1}\bar{q}_{2}\right\rangle }_{1},\quad \chi
_{2}^{M}={\left\vert q_{1}\bar{q}_{2}\right\rangle }_{0},  \label{spinM}
\end{equation}%
where subscript $J=0$ or $1$ outside the bracket denotes the total spin of
hadron. The spin-color wavefunctions with spin $J=0$ and $1$ are then
\begin{equation}
\phi ^{M}\chi _{1}^{M}={\left\vert q_{1}\bar{q}_{2}\right\rangle }_{1},\quad
\phi ^{M}\chi _{2}^{M}={\left\vert q_{1}\bar{q}_{2}\right\rangle }_{0}.
\label{colorspinM}
\end{equation}

\textbf{Baryons}: The color wavefunctions $\phi ^{B}=\left\vert {\left(
q_{1}q_{2}\right) }^{\bar{3}}q_{3}\right\rangle $ can be in one of three
spin states
\begin{equation}
\begin{aligned} \chi_{1}^{B}={\left| {\left(q_{1}q_{2}\right)}_{1} q_{3}
\right\rangle}_{3/2}, \\ \chi_{2}^{B}={\left| {\left(q_{1}q_{2}\right)}_{1}
q_{3} \right\rangle}_{1/2}, \\ \chi_{3}^{B}={\left|
{\left(q_{1}q_{2}\right)}_{0} q_{3} \right\rangle}_{1/2}, \end{aligned}
\label{spinB}
\end{equation}%
where $\left( q_{1}q_{2}\right) $ stands for a diquark with spin $J=1$ or $0$
in color configuration $\boldsymbol{\bar{3}}_{c}$.

To write wavefunction for a hadron, the flavor symmetry has to be
considered. For a flavor-symmetric wavefunction of $\left( q_{1}q_{2}\right)
$, with isospin $I=1$ or identical flavors, we use a symbol $\delta
_{12}^{S}=1$. For a flavor-asymmetric wavefunction with $I=0$, a symbol $%
\delta _{12}^{A}=1$ will be used. For two quarks $q_{1}$ and $q_{2}$ with
different flavors which goes beyond isospin symmetry, one can use $\delta
_{12}^{S}=\delta _{12}^{A}=1$. With the help of Pauli principle, one can
write three flavor-spin-color wavefunctions for baryons
\begin{equation}
\begin{aligned} \phi^{B}\chi_{1}^{B}={\left|
{\left(q_{1}q_{2}\right)}_{1}^{\bar{3}} q_{3} \right\rangle}_{3/2}
\delta_{12}^{S}, \\ \phi^{B}\chi_{2}^{B}={\left|
{\left(q_{1}q_{2}\right)}_{1}^{\bar{3}} q_{3} \right\rangle}_{1/2}
\delta_{12}^{S}, \\ \phi^{B}\chi_{3}^{B}={\left|
{\left(q_{1}q_{2}\right)}_{0}^{\bar{3}} q_{3} \right\rangle}_{1/2}
\delta_{12}^{A}. \end{aligned}  \label{colorspinB}
\end{equation}

Owing to the non-diagonal chromomagnetic interaction (\ref{CMI}), some
hadronic states with same $J^{P}$ but different spin-color wavefunctions can
mix(CMI mixing). For example, for the doubly heavy baryons $\Xi _{bc}$ and $%
\Xi _{bc}^{\prime }$ with flavor structure $(bc)s$ the use of $\delta
_{12}^{S}=\delta _{12}^{A}=1$ is not enough to distinguish the two
configurations $\phi ^{B}\chi _{2}^{B}$ and $\phi ^{B}\chi _{3}^{B}$ solely
in terms of their $J^{P}$ quantum numbers. Thus, the physical state must be
one of the mixing states of them. See Sect. IV for the details
of chromomagnetic mixing.

\textbf{Tetraquarks}: A tetraquark can have the color structure of whether $%
\boldsymbol{6}_{c}\otimes \boldsymbol{\bar{6}}_{c}$ or $\boldsymbol{\bar{3}}_{c}\otimes
\boldsymbol{3}_{c}$, with the respective color wavefunctions,
\begin{equation}
\phi _{1}^{T}=\left\vert {\left( q_{1}q_{2}\right) }^{6}{\left( \bar{q}_{3}%
\bar{q}_{4}\right) }^{\bar{6}}\right\rangle ,\quad \phi _{2}^{T}=\left\vert {%
\left( q_{1}q_{2}\right) }^{\bar{3}}{\left( \bar{q}_{3}\bar{q}_{4}\right) }%
^{3}\right\rangle ,  \label{colorT}
\end{equation}%
and it can be one of the following six states
\begin{equation}
\begin{aligned} \chi_{1}^{T}={\left| {\left(q_{1}q_{2}\right)}_{1}
{\left(\bar{q}_{3}\bar{q}_{4}\right)}_{1} \right\rangle}_{2}, \quad
\chi_{2}^{T}={\left| {\left(q_{1}q_{2}\right)}_{1}
{\left(\bar{q}_{3}\bar{q}_{4}\right)}_{1} \right\rangle}_{1}, \\
\chi_{3}^{T}={\left| {\left(q_{1}q_{2}\right)}_{1}
{\left(\bar{q}_{3}\bar{q}_{4}\right)}_{1} \right\rangle}_{0}, \quad
\chi_{4}^{T}={\left| {\left(q_{1}q_{2}\right)}_{1}
{\left(\bar{q}_{3}\bar{q}_{4}\right)}_{0} \right\rangle}_{1}, \\
\chi_{5}^{T}={\left| {\left(q_{1}q_{2}\right)}_{0}
{\left(\bar{q}_{3}\bar{q}_{4}\right)}_{1} \right\rangle}_{1}, \quad
\chi_{6}^{T}={\left| {\left(q_{1}q_{2}\right)}_{0}
{\left(\bar{q}_{3}\bar{q}_{4}\right)}_{0} \right\rangle}_{0}, \end{aligned}
\label{spinT}
\end{equation}%
which lead to twelve basis wavefunctions
\begin{equation}
\begin{aligned} \phi_{1}^{T}\chi_{1}^{T}={\left|
{\left(q_{1}q_{2}\right)}_{1}^{6}
{\left(\bar{q}_{3}\bar{q}_{4}\right)}_{1}^{\bar{6}} \right\rangle}_{2}
\delta_{12}^{A}\delta_{34}^{A}, \\ \phi_{2}^{T}\chi_{1}^{T}={\left|
{\left(q_{1}q_{2}\right)}_{1}^{\bar{3}}
{\left(\bar{q}_{3}\bar{q}_{4}\right)}_{1}^{3} \right\rangle}_{2}
\delta_{12}^{S}\delta_{34}^{S}, \\ \phi_{1}^{T}\chi_{2}^{T}={\left|
{\left(q_{1}q_{2}\right)}_{1}^{6}
{\left(\bar{q}_{3}\bar{q}_{4}\right)}_{1}^{\bar{6}} \right\rangle}_{1}
\delta_{12}^{A}\delta_{34}^{A}, \\ \phi_{2}^{T}\chi_{2}^{T}={\left|
{\left(q_{1}q_{2}\right)}_{1}^{\bar{3}}
{\left(\bar{q}_{3}\bar{q}_{4}\right)}_{1}^{3} \right\rangle}_{1}
\delta_{12}^{S}\delta_{34}^{S}, \\ \phi_{1}^{T}\chi_{3}^{T}={\left|
{\left(q_{1}q_{2}\right)}_{1}^{6}
{\left(\bar{q}_{3}\bar{q}_{4}\right)}_{1}^{\bar{6}} \right\rangle}_{0}
\delta_{12}^{A}\delta_{34}^{A}, \\ \phi_{2}^{T}\chi_{3}^{T}={\left|
{\left(q_{1}q_{2}\right)}_{1}^{\bar{3}}
{\left(\bar{q}_{3}\bar{q}_{4}\right)}_{1}^{3} \right\rangle}_{0}
\delta_{12}^{S}\delta_{34}^{S}, \\ \phi_{1}^{T}\chi_{4}^{T}={\left|
{\left(q_{1}q_{2}\right)}_{1}^{6}
{\left(\bar{q}_{3}\bar{q}_{4}\right)}_{0}^{\bar{6}} \right\rangle}_{1}
\delta_{12}^{A}\delta_{34}^{S}, \\ \phi_{2}^{T}\chi_{4}^{T}={\left|
{\left(q_{1}q_{2}\right)}_{1}^{\bar{3}}
{\left(\bar{q}_{3}\bar{q}_{4}\right)}_{0}^{3} \right\rangle}_{1}
\delta_{12}^{S}\delta_{34}^{A}, \\ \phi_{1}^{T}\chi_{5}^{T}={\left|
{\left(q_{1}q_{2}\right)}_{0}^{6}
{\left(\bar{q}_{3}\bar{q}_{4}\right)}_{1}^{\bar{6}} \right\rangle}_{1}
\delta_{12}^{S}\delta_{34}^{A}, \\ \phi_{2}^{T}\chi_{5}^{T}={\left|
{\left(q_{1}q_{2}\right)}_{0}^{\bar{3}}
{\left(\bar{q}_{3}\bar{q}_{4}\right)}_{1}^{3} \right\rangle}_{1}
\delta_{12}^{A}\delta_{34}^{S}, \\ \phi_{1}^{T}\chi_{6}^{T}={\left|
{\left(q_{1}q_{2}\right)}_{0}^{6}
{\left(\bar{q}_{3}\bar{q}_{4}\right)}_{0}^{\bar{6}} \right\rangle}_{0}
\delta_{12}^{S}\delta_{34}^{S}, \\ \phi_{2}^{T}\chi_{6}^{T}={\left|
{\left(q_{1}q_{2}\right)}_{0}^{\bar{3}}
{\left(\bar{q}_{3}\bar{q}_{4}\right)}_{0}^{3} \right\rangle}_{0}
\delta_{12}^{A}\delta_{34}^{A}. \end{aligned}  \label{colorspinT}
\end{equation}

We list all relevant color wavefunctions in Appendix A and spin
wavefunctions in Appendix B. With them, one can evaluate the color and spin
factors in Eq.~(\ref{CMI}) with the help of the following formulas
\begin{equation}
{\left\langle \boldsymbol{\lambda _{i}}\cdot \boldsymbol{\lambda _{j}}%
\right\rangle }_{nm}=\sum_{\alpha =1}^{8}Tr\left( c_{in}^{\dagger }\lambda
^{\alpha }c_{im}\right) Tr\left( c_{jn}^{\dagger }\lambda ^{\alpha
}c_{jm}\right) ,  \label{colorfc}
\end{equation}%
\begin{equation}
{\left\langle \boldsymbol{\sigma _{i}}\cdot \boldsymbol{\sigma _{j}}%
\right\rangle }_{xy}=\sum_{\alpha =1}^{3}Tr\left( \chi _{ix}^{\dagger
}\sigma ^{\alpha }\chi _{iy}\right) Tr\left( \chi _{jx}^{\dagger }\sigma
^{\alpha }\chi _{jy}\right) ,  \label{spinfc}
\end{equation}%
where $n$, $m$ and $x$, $y$ indicate the specific color and spin states
respectively, $i$ and $j$ are the indexes of quarks (antiquarks), and the
functions $c$ and $\chi $ are the respective basis vectors in the color and
spin spaces. Table \ref{tab:CMI} lists a set of non-mixed hadronic states
with their respective CMI's.

Now, we are in the position to construct the matrix formula of CMI energy (%
\ref{CMI}) and diagonalize it so as to minimize the obtained mass formula.
Adding the binding energy (for heavy quark pair and for a pair of one heavy
quark and one strange quark) to the bag energy, one can solve the dynamical
parameters $x_{i}$ and $R$, and thereby obtain the wavefunctions of a given
hadron.

\renewcommand{\tabcolsep}{0.20cm} \renewcommand{\arraystretch}{1.2}
\begin{table*}[!htb]
\caption{Chromomagnetic interactions for the non-mixing hadrons with
respective wavefunctions. $C_{ij}$ follows Eq.~(\protect\ref{Cij}) with
subscripts corresponding to quark or antiquark.}
\label{tab:CMI}%
\begin{tabular}{cccccc}
\hline\hline
\textrm{State} & \textrm{Wave\ Function} & \textrm{CMI} & \textrm{State} &
\textrm{Wave\ Function} & \textrm{CMI} \\ \hline
$\pi$ & $\phi^{M}\chi_{2}^{M}$ & $-16C_{nn}$ & $\omega$ & $%
\phi^{M}\chi_{1}^{M}$ & $\frac{16}{3}C_{nn}$ \\
$K$ & $\phi^{M}\chi_{2}^{M}$ & $-16C_{sn}$ & $K^{\ast}$ & $%
\phi^{M}\chi_{1}^{M}$ & $\frac{16}{3}C_{sn}$ \\
&  &  & $\phi$ & $\phi^{M}\chi_{1}^{M}$ & $\frac{16}{3}C_{ss}$ \\
$D$ & $\phi^{M}\chi_{2}^{M}$ & $-16C_{cn}$ & $D^{\ast}$ & $%
\phi^{M}\chi_{1}^{M}$ & $\frac{16}{3}C_{cn}$ \\
$D_{s}$ & $\phi^{M}\chi_{2}^{M}$ & $-16C_{cs}$ & $D_{s}^{\ast}$ & $%
\phi^{M}\chi_{1}^{M}$ & $\frac{16}{3}C_{cs}$ \\
$\eta_{c}$ & $\phi^{M}\chi_{2}^{M}$ & $-16C_{cc}$ & $J/\psi$ & $%
\phi^{M}\chi_{1}^{M}$ & $\frac{16}{3}C_{cc}$ \\
$N$ & $\phi^{B}\chi_{3}^{B}$ & $-8C_{nn}$ & $\Delta$ & $\phi^{B}\chi_{1}^{B}$
& $8C_{nn}$ \\
$\Lambda$ & $\phi^{B}\chi_{3}^{B}$ & $-8C_{nn}$ &  &  &  \\
$\Sigma$ & $\phi^{B}\chi_{2}^{B}$ & $\frac{8}{3}C_{nn}-\frac{32}{3}C_{sn}$ &
$\Sigma^{\ast}$ & $\phi^{B}\chi_{1}^{B}$ & $\frac{8}{3}C_{nn}+\frac{16}{3}%
C_{sn}$ \\
$\Lambda_{c}$ & $\phi^{B}\chi_{3}^{B}$ & $-8C_{nn}$ &  &  &  \\
$\Sigma_{c}$ & $\phi^{B}\chi_{2}^{B}$ & $\frac{8}{3}C_{nn}-\frac{32}{3}%
C_{cn} $ & $\Sigma_{c}^{\ast}$ & $\phi^{B}\chi_{1}^{B}$ & $\frac{8}{3}C_{nn}+%
\frac{16}{3}C_{cn}$ \\
$\Xi$ & $\phi^{B}\chi_{2}^{B}$ & $\frac{8}{3}C_{ss}-\frac{32}{3}C_{sn}$ & $%
\Xi^{\ast}$ & $\phi^{B}\chi_{1}^{B}$ & $\frac{8}{3}C_{ss}+\frac{16}{3}C_{sn}$
\\
$\Xi_{c}$ & $\phi^{B}\chi_{3}^{B}$ & $-8C_{sn}$ & & & \\
$\Xi_{c}^{\prime}$ & $\phi^{B}\chi_{2}^{B}$ & $\frac{8}{3}C_{sn}-\frac{16}{3}C_{cn}-\frac{16}{3}C_{cs}$
& $\Xi_{c}^{\ast}$ & $\phi^{B}\chi_{1}^{B}$ & $\frac{8}{3}C_{sn}+\frac{8}{3}C_{cn}+\frac{8}{3}%
C_{cs}$ \\
&  &  & $\Omega$ & $\phi^{B}\chi_{1}^{B}$ & $8C_{ss}$ \\
$ss\bar{c}\bar{c}$ & $\phi_{2}^{T}\chi_{2}^{T}$ & $\frac{8}{3} C_{ss}-%
\frac{16}{3}C_{cs}+\frac{8}{3}C_{cc}$ & $ss\bar{c}\bar{c}$ & $\phi_{2}^{T}%
\chi_{1}^{T}$ & $\frac{8}{3}C_{ss}+\frac{16}{3}C_{cs}+\frac{8}{3}C_{cc}$ \\
${\left(nn\bar{c}\bar{c}\right)}^{I=1}$ & $\phi_{2}^{T}\chi_{2}^{T}$ & $%
\frac{8}{3}C_{nn}-\frac{16}{3}C_{cn}+\frac{8}{3}C_{cc}$ & ${\left(nn\bar{c}%
\bar{c}\right)}^{I=1}$ & $\phi_{2}^{T}\chi_{1}^{T}$ & $\frac{8}{3}C_{nn}+%
\frac{16}{3}C_{cn}+\frac{8}{3}C_{cc}$ \\
\hline\hline
\end{tabular}%
\end{table*}

\renewcommand{\tabcolsep}{0.41cm} \renewcommand{\arraystretch}{1.2}
\begin{table*}[!htb]
\caption{Sum rule for magnetic moments of spin states of
mesons $\left(q_{1}{\bar{q}}_{2}\right)$, baryons $\left(q_{1}q_{2}\right)q_{3}$,
and tetraquarks $\left(q_{1}q_{2}\right)\left({\bar{q}}_{3}{\bar{q}}_{4}\right)$
and their spin-mixed systems.}
\label{tab:musum}%
\begin{tabular}{cc}
\hline\hline
$\psi_{spin}$ & $\mu$ \\ \hline
$\chi_{1}^{M}$ & $\mu_{1}+\mu_{2}$ \\
$\chi_{2}^{M}$ & 0 \\
$\chi_{1}^{B}$ & $\mu_{1}+\mu_{2}+\mu_{3}$
\\
$\chi_{2}^{B}$ & $\frac{1}{3}\left(2\mu_{1}+2\mu_{2}-\mu_{3}\right)$ \\
$\chi_{3}^{B}$ & $\mu_{3}$ \\
$C_{1}\chi_{2}^{B}+C_{2}\chi_{3}^{B}$ & $C_{1}^{2}\mu\left(\chi_{2}^{B}%
\right) + C_{2}^{2}\mu\left(\chi_{3}^{B}\right) + \frac{2}{\sqrt{3}}%
C_{1}C_{2}\left(\mu_{2}-\mu_{1}\right)$ \\
$\chi_{1}^{T}$ & $\mu_{1}+\mu_{2}+\mu_{3}+\mu_{4}$ \\
$\chi_{2}^{T}$ & $\frac{1}{2}\left(\mu_{1}+\mu_{2}+\mu_{3}+\mu_{4}\right)$
\\
$\chi_{3}^{T}$ & 0 \\
$\chi_{4}^{T}$ & $\mu_{1}+\mu_{2}$ \\
$\chi_{5}^{T}$ & $\mu_{3}+\mu_{4}$ \\
$\chi_{6}^{T}$ & 0 \\
$C_{1}\chi_{3}^{T}+C_{2}\chi_{6}^{T}$ & 0 \\
$C_{1}\chi_{5}^{T}+C_{2}\chi_{4}^{T}$ & $C_{1}^{2}\mu\left(\chi_{5}^{T}%
\right) + C_{2}^{2}\mu\left(\chi_{4}^{T}\right)$ \\
$C_{1}\chi_{2}^{T}+C_{2}\chi_{4}^{T}+C_{3}\chi_{5}^{T}$ & $%
C_{1}^{2}\mu\left(\chi_{2}^{T}\right) +
C_{2}^{2}\mu\left(\chi_{4}^{T}\right) +
C_{3}^{2}\mu\left(\chi_{5}^{T}\right) + \sqrt{2}C_{1}C_{2}\left(\mu_{3}-%
\mu_{4}\right) + \sqrt{2}C_{1}C_{3}\left(\mu_{2}-\mu_{1}\right)$ \\
$C_{1}\chi_{2}^{T}+C_{2}\chi_{5}^{T}+C_{3}\chi_{4}^{T}$ & $%
C_{1}^{2}\mu\left(\chi_{2}^{T}\right) +
C_{2}^{2}\mu\left(\chi_{5}^{T}\right) +
C_{3}^{2}\mu\left(\chi_{4}^{T}\right) + \sqrt{2}C_{1}C_{2}\left(\mu_{2}-%
\mu_{1}\right) + \sqrt{2}C_{1}C_{3}\left(\mu_{3}-\mu_{4}\right)$ \\
\hline\hline
\end{tabular}%
\end{table*}

\subsection{Hadronic Properties}

Given the parameters $x_{i}$ and $R$ describing a hadronic state, mass and
other properties(e.g.,the charge radius and magnetic moment) can be
evaluated. Following the standard method, one can firstly calculate the
contribution of a quark or an antiquark $i$ with electric charge $Q_{i}$ to
charge radius \cite{DeGrand:1975cf}
\begin{equation}
\begin{aligned} {\left\langle r_{E}^{2} \right\rangle}_{i}&= Q_{i}R^{2}
\frac{\alpha_{i}
\left[2x_{i}^{2}\left(\alpha_{i}-1\right)+4\alpha_{i}+2\lambda_{i}-3\right]}
{3x_{i}^{2} \left[2\alpha_{i}\left(\alpha_{i}-1\right)+\lambda_{i}\right]}
\\ &-Q_{i}R^{2} \frac{\lambda_{i}
\left[4\alpha_{i}+2\lambda_{i}-2x_{i}^{2}-3\right]} {2x_{i}^{2}
\left[2\alpha_{i}\left(\alpha_{i}-1\right)+\lambda_{i}\right]}. \end{aligned}
\label{rEi}
\end{equation}%
The sum of Eq.~(\ref{rEi}) then gives the charge radius of a hadronic state
\cite{Chodos:1974pn}
\begin{equation}
r_{E}={\left\vert \sum\nolimits_{i}{\left\langle r_{E}^{2}\right\rangle }%
_{i}\right\vert }^{1/2}.  \label{rEsum}
\end{equation}%
We note that Eq.~(\ref{rEsum}) also holds true for the chromomagnetic-mixing
systems having the identical quark constituents.

For magnetic moment, the following equations \cite%
{DeGrand:1975cf,Wang:2016dzu}, which are computed relative to the magnetic moment of proton and has the unit of $\mu_{p}$, are useful:
\begin{equation}
\mu _{i}=Q_{i}\bar{\mu}_{i}=Q_{i}\frac{R}{6}\frac{4\alpha _{i}+2\lambda
_{i}-3}{2\alpha _{i}\left( \alpha _{i}-1\right) +\lambda _{i}},  \label{mui}
\end{equation}%
\begin{equation}
\mu =\left\langle \psi _{spin}\left\vert \sum\nolimits_{i}g_{i}\mu
_{i}S_{iz}\right\vert \psi _{spin}\right\rangle ,  \label{musum}
\end{equation}%
where $g_{i}=2$, and $S_{iz}$ is the third component of spin for an
individual quark or antiquark. In all Tables for the results of
magnetic moments, obtained from Eq. (23) and Eq. (24), we transform them into that in the unit of the nuclear magneton $\mu_{N}$, with the help of the measured data $\mu_{p}=2.79285\mu _{N}$. If the chromomagnetic mixing enters, the
total spin wavefunction becomes
\begin{equation}
\left\vert \psi _{spin}\right\rangle =C_{1}\chi _{1}+C_{2}\chi _{2},
\end{equation}%
by which Eq. (\ref{musum}) gives
\begin{equation}
\mu =C_{1}^{2}\mu \left( \chi _{1}\right) +C_{2}^{2}\mu \left( \chi
_{2}\right) +2C_{1}C_{2}\mu ^{\mathrm{tr}}\left( \chi _{1},\chi _{2}\right) ,
\end{equation}%
with $\mu ^{\mathrm{tr}}$ the cross-term standing for transition moment \cite%
{Bernotas:2012nz} and $(C_{1},C_{2})$ the eigenvector of the given mixing
state. We list all spin wavefunctions in Appendix B, and derive magnetic
moments for them and their possibly-mixed systems involved in this work. The
results for the spin wavefunctions and the respective magnetic moments are
listed in Table \ref{tab:musum} collectively.

Note that the cross-terms in CMI-mixing systems are not symmetric under the
exchange between quarks $q_{1}$ and $q_{2}$ or, between $q_{3}$ and $q_{4}$
in the flavor space. While the expression of cross-term for the diquark $%
(ud) $ differs a sign for $(ud)$ and $(du)$ within the symmetric or
asymmetric flavor wavefunctions when $I_{3}=1$ or -1 in isospin space,
respectively, the explicit computation via these wavefunctions can offset
such cross-term. Similar conclusions also apply as the hadron systems
respect the $SU(2)$ isospin symmetry.

\section{Determination of Parameters}

\label{Parameters} In MIT bag model, the parameters (nonstrange $m_{n}$ and
strange $m_{s}$ quark masses, $B$, $Z_{0}$ and $\alpha _{s}$) are determined
based on the mass spectra of the light hadrons $N$, $\Delta $, $\omega $ and
$\Omega $ in their ground states. The results read \cite{DeGrand:1975cf}
\begin{equation}
\begin{Bmatrix}
m_{n}=0, & m_{s}=0.279\,\text{GeV,} &  \\
Z_{0}=1.83, & B^{1/4}=0.145\,\text{GeV,} & \alpha _{s}=0.55.%
\end{Bmatrix}
\label{oripar}
\end{equation}

We choose Eq.~(\ref{oripar}) to be the parameters applying to both of light and
heavy hadrons, with one exception that the strong coupling $\alpha _{s}$
changes with the size $R$ of hadron around $0.55$, as given by Eq. (\ref%
{alphaS-mine}). To fix the model parameters it remains two tasks yet.

The first task is to extract the heavy quark masses $m_{c}$ and $m_{b}$.
Given Eq. (\ref{oripar}), one can apply Eqs. (\ref{MBm}) and (\ref{CMI}) to
the heavy-light mesons $D^{\ast }$ and $B^{\ast }$ to fix numerically $m_{c}$
and $m_{b}$, respectively. The results are
\begin{equation}
\begin{Bmatrix}
m_{c}=1.641\,\text{GeV}, & m_{b}=5.093\,\text{GeV}%
\end{Bmatrix}%
.  \label{mcmb}
\end{equation}

The second is to fix the binding energy $B_{QQ^{\prime }}$(and $B_{Q\bar{Q}%
^{\prime }}$, $Q,Q^{\prime }=s,c,b$ here), which is proposed in Ref. \cite%
{KR:2014gca} to occur in charmed-strange hadrons, bottom-strange hadrons and
heavy quadrennia. It can be due to nontrivial short-range interaction
between two heavy quarks and between heavy and strange quarks\cite%
{KR:2014gca,Karliner:2017elp,Karliner:2018bms}. Applied to the strange heavy
mesons($Q\bar{s}$ and $Q\bar{Q\prime }$), this binding enters the mass
formula through
\begin{eqnarray}
M(Q\bar{s}) &=&\omega _{Q}+\omega _{s}+\frac{4}{3}\pi R^{3}B-\frac{Z_{0}}{R}%
+\langle H_{CMI}\rangle +B_{Q\bar{s}},  \notag \\
M(Q\bar{Q\prime }) &=&\omega _{Q}+\omega _{Q\prime }+\frac{4}{3}\pi R^{3}B-%
\frac{Z_{0}}{R}+\langle H_{CMI}\rangle +B_{Q\bar{Q\prime }}.  \label{HLm0}
\end{eqnarray}

Applying to the case of the vector mesons $D_{s}^{\ast }=c\bar{s}$, this
allows one to solve the binding term,
\begin{equation}
B_{c\bar{s}}=M(D_{s}^{\ast })-\omega _{c}-\omega _{s}-\frac{4}{3}\pi R^{3}B+%
\frac{Z_{0}}{R}-\langle H_{CMI}\rangle ,  \label{BB0}
\end{equation}%
with the bag radius $R$ solved variationally for the $D_{s}^{\ast }$ mesons.
Numerically, one finds, $B_{c\bar{s}}=-0.050\,$GeV by Eq. (\ref{BB0}). Note
that the short-range color interaction of the quark pair $c\bar{s}$ in color
singlet($1_{c}$) in heavy meson $D_{s}^{\ast }$ can be related to that of
the $cs$ pair in color antitriplet($\bar{3}_{c}$) in a heavy baryon $nsc$ by
the factor $1/2$, one can reasonably assume, in the short range, that the
strength of the $cs$ interaction in $\bar{3}_{c}$ is half that of $c\bar{s}$
in $1_{c}$. This follows that $B_{cs}=B_{c\bar{s}}/2=-0.025\,$GeV. Here, the
factor half can be extracted from the ratio of the color factor $-8/3$ in
Eq. (\ref{cfcM}) for $1_{c}$ and $-16/3$ Eq. (\ref{cfcB}), evaluated in
Appendix A:$1/2=(-8/3)/(-16/3)$. The same holds true also for the quark pair
$Q\bar{Q\prime }$ in the heavy mesons $B_{s}^{\ast }$, $J/\psi $, $\Upsilon $
and $B_{c}^{\ast }$. Thus, for the quark pair($Qs$ an $QQ\prime $) in heavy
baryon, we choose
\begin{equation}
B_{Qs}=B_{Q\bar{s}}/2,B_{QQ\prime }=B_{Q\bar{Q\prime }}/2,  \label{half}
\end{equation}%
for the quark pair in $\boldsymbol{\bar{3}_{c}}$ and the quark-antiquark
pair in $1_{c}$, and solve the model (\ref{MBm}) for the heavy mesons $%
B_{s}^{\ast }$, $J/\psi $, $\Upsilon $ and $B_{c}^{\ast }$, obtaining, by
Eqs. (\ref{HLm0}) and (\ref{half}),
\begin{equation}
\begin{Bmatrix}
B_{cs}=B_{c\bar{s}}/2=-0.025\,\text{GeV,} & B_{cc}=B_{c\bar{c}}/2=-0.077\,%
\text{GeV,} \\
B_{bs}=B_{b\bar{s}}/2=-0.032\,\text{GeV,} & B_{bb}=B_{b\bar{b}}/2=-0.128\,%
\text{GeV,} \\
B_{bc}=B_{b\bar{c}}/2=-0.101\,\text{GeV,} &
\end{Bmatrix}
\label{Bcs}
\end{equation}%
where the results for the $c\bar{s}$ pair is also included. Here in
computation, we have used the mass $M(B_{c}^{\ast })=6.332$ GeV of the heavy
meson $B_{c}^{\ast }$ in Ref.~\cite{Ebert:2002pp}, due to lacking of the
measured $B_{c}^{\ast }$.

\vspace{-0.6cm}

\begin{figure}[!th]
\setlength{\abovecaptionskip}{-1.0cm}
\begin{center}
\includegraphics[width=0.77\textwidth]{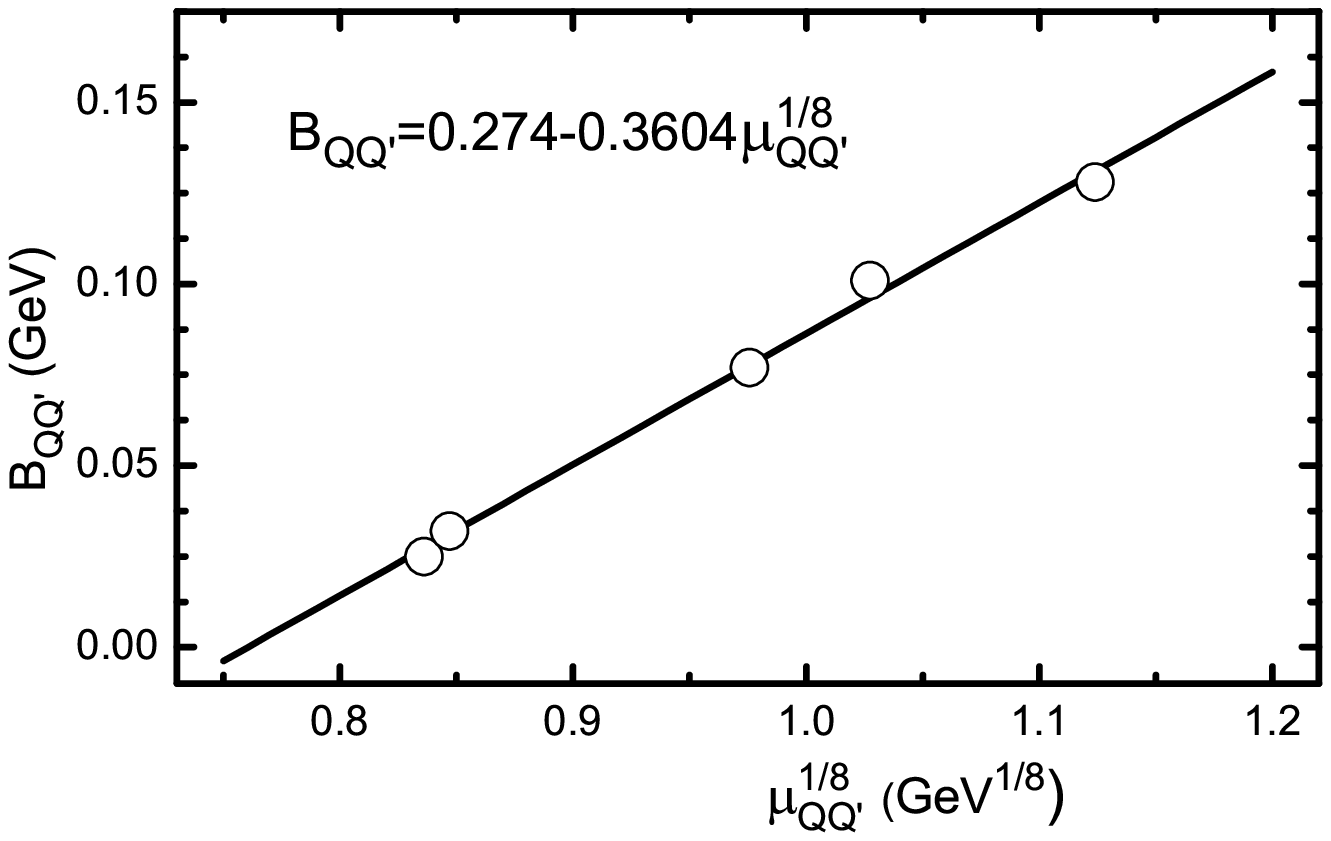}
\end{center}
\caption{ Binding energy $B_{QQ^{\prime }}$ (solid line) in Eq. (\protect\ref%
{Bcs}) as a function of the reduced mass $\protect\mu _{QQ^{\prime }}$ of
two quarks $Q$ and $Q^{\prime }$. Circles correspond to the pair data {($%
\protect\mu _{QQ^{\prime }}$,$B_{QQ^{\prime }}$)} with the respective $%
QQ^{\prime }$=$sc,sb,cc,cb,bb$. }
\label{fig:binding}
\end{figure}

\setlength{\abovecaptionskip}{-0.8cm}

It can be seen from Eq.~(\ref{Bcs}) that $B_{QQ^{\prime }}$ depends
monotonically on the reduced mass $\mu _{QQ^{\prime }}=m_{Q}m_{Q^{\prime
}}/(m_{Q}+m_{Q^{\prime }})$ of two involved quarks $Q$ and $Q^{\prime }$.
The dependence (FIG.~\ref{fig:binding}) can be approximated by

\begin{equation}
B_{QQ^{\prime }}\text{(}\bar{3}_{c}\text{)}=0.274\,\text{GeV}-0.3604(\text{%
GeV}^{7/8})\mu _{QQ^{\prime }}^{1/8}\text{.}  \label{BQQ}
\end{equation}

\renewcommand{\tabcolsep}{0.24cm} \renewcommand{\arraystretch}{1.1}
\begin{table}[!htb]
\setlength{\abovecaptionskip}{0.4cm}
\caption{Computed masses (in GeV), magnetic moments (in $\protect\mu_{N}$)
and charge radii (in fm) of light ground-states mesons, compared to the measured data.
The blank cells indicate the values same with the above .}
\label{tab:lightmeson}%
\begin{tabular}{ccccccc}
\hline\hline
\textrm{State} & $R_{0}$(GeV$^{-1}$) & $M_{bag}$ & $M_{exp}$ \cite{Tanabashi:D18}
& $\mu_{bag}$ & $r_{E}$ (fm) & $r_{E}$ (fm)\cite{Tanabashi:D18} \\ \hline
$\pi^{+}$ & 4.31 & 0.348 & 0.140 & - & 0.62 & 0.66 \\
$\omega$ & 4.55 & 0.776 & 0.783 & 0 & 0 & - \\
$K^{+}$ & 4.34 & 0.561 & 0.494 & - & 0.61 & 0.56 \\
$K^{0}$ &  &  & 0.498 & - & 0.13 & 0.28 \\
$K^{\ast +}$ & 4.63 & 0.918 & 0.892 & 2.30 & 0.65 & - \\
$K^{\ast 0}$ &  &  & 0.896 & -0.18 & 0.15 & - \\
$\phi$ & 4.70 & 1.064 & 1.019 & 0 & 0 & - \\ \hline\hline
\end{tabular}%
\end{table}

\renewcommand{\tabcolsep}{0.3cm} \renewcommand{\arraystretch}{1.1}
\begin{table}[!htb]
\caption{Computed masses(in GeV), magnetic moments (in $\protect\mu_{N}$)
and charge radii (in fm) of heavy mesons in their ground-states. The blank
cell follows $5.325\,\mathrm{GeV}$ indicates the values same with the above.}
\label{tab:heavymeson}%
\begin{tabular}{cccccc}
\hline\hline
\textrm{State} & $R_{0}$(GeV$^{-1}$) & $M_{bag}$ & $M_{exp}$ \cite{Tanabashi:D18}
& $\mu_{bag}$ & $r_{E}$(fm) \\ \hline
$D^{+}$ & 3.63 & 1.835 & 1.870 & - & 0.46 \\
$D^{0}$ &  &  & 1.865 & - & 0.25 \\
$D^{\ast +}$ & 4.09 & 2.009[input] & 2.010 & 1.21 & 0.51 \\
$D^{\ast 0}$ &  &  & 2.007 & -0.98 & 0.29 \\
$B^{+}$ & 3.14 & 5.248 & 5.279 & - & 0.42 \\
$B^{0}$ &  &  & 5.280 & - & 0.17 \\
$B^{\ast +}$ & 3.47 & 5.325[input] & 5.325 & 1.32 & 0.46 \\
$B^{\ast 0}$ &  &  &  & -0.53 & 0.19 \\
$D_{s}^{+}$ & 3.77 & 1.961 & 1.968 & - & 0.46 \\
$D_{s}^{\ast +}$ & 4.17 & 2.112[input] & 2.112 & 1.08 & 0.51 \\
$B_{s}^{0}$ & 3.35 & 5.346 & 5.367 & - & 0.16 \\
$B_{s}^{\ast 0}$ & 3.62 & 5.415[input] & 5.415 & 1.01 & 0.17 \\
$\eta_{c}$ & 3.15 & 3.002 & 2.984 & - & 0 \\
$J/\psi$ & 3.54 & 3.097[input] & 3.097 & 0 & 0 \\
$B_{c}^{+}$ & 2.53 & 6.273 & 6.274 & - & 0.29 \\
$B_{c}^{\ast +}$ & 2.81 & 6.332[input] & 6.332 & 0.52 & 0.32 \\
$\eta_{b}$ & 1.59 & 9.396 & 9.399 & - & 0 \\
$\Upsilon$ & 1.80 & 9.460[input] & 9.460 & 0 & 0 \\ \hline\hline
\end{tabular}%
\end{table}

One can scale Eq. (\ref{Bcs}) for the pair in $\boldsymbol{\bar{3}}_{c}$ to
the pair $QQ^{\prime }$ in other color configurations. This can be done by
computing the explicit ratios of the color factors in Eq.~(\ref{colorfc})
(evaluated in Appendix A). The scale factor $g([QQ^{\prime }]_{R})$($=$ratio
of the color factor for representation $R$ and the color factor for $\bar{3}%
_{c}$) for pair $QQ^{\prime }$($=bb,cc,bc,bs,cs$) can be given explicitly by
\begin{equation}
\left\{
\begin{array}{cc}
g([QQ^{\prime }]_{\boldsymbol{1}_{c}})= & 2, \\
g([QQ^{\prime }]_{\boldsymbol{6}_{c}})= & -1/2,%
\end{array}%
\begin{array}{cc}
g([b\bar{s}]_{\boldsymbol{6}_{c}})=g([c\bar{s}]_{\boldsymbol{6}_{c}})= & 5/4,
\\
g([b\bar{s}]_{\boldsymbol{\bar{3}}_{c}})=g([c\bar{s}]_{\boldsymbol{\bar{3}}_{c}})= & 1/2,%
\end{array}%
\right\}  \label{gf}
\end{equation}%
and the binding energy between the pair $QQ^{\prime }$ in $R$ is then
\begin{equation}
B([QQ^{\prime }]_{R})=g([QQ^{\prime }]_{R})B_{QQ^{\prime }}\text{.}
\label{BQQR}
\end{equation}%
with $B_{QQ^{\prime }}\equiv B([QQ^{\prime }]_{\boldsymbol{\bar{3}}_{c}})$
given in Eq. (\ref{Bcs}). For instance, for a pair $QQ^{\prime }$ in $R=%
\boldsymbol{1}_{c}$, the scaled result for the binding energy $B([QQ^{\prime
}]_{\boldsymbol{1}_{c}})=2B_{QQ^{\prime }}$. For a pair in $R=\boldsymbol{6}%
_{c}$, it is $-B_{QQ^{\prime }}/2$. For $Q\bar{s}$ in tetraquark $\bar{q}%
\bar{s}QQ^{\prime }$, the scaled binding energy is $5B_{QQ^{\prime }}/4$ for
$Q\bar{s}$ in $\boldsymbol{6}_{c}\otimes \boldsymbol{\bar{6}}_{c}$, and is $%
B_{QQ^{\prime }}/2$ for $Q\bar{s}$ in $\boldsymbol{3}_{c}\otimes \boldsymbol{%
\bar{3}}_{c}$. The total binding energy of the baryons and tetraquark
systems are given by the sum of all pair binding energies and can be found
in Eqs.~(\ref{phbinding}),(\ref{ph1binding}),(\ref{ph2binding}) in Appendix
C .

\renewcommand{\tabcolsep}{0.25cm} \renewcommand{\arraystretch}{1.1}
\begin{table}[!htb]
\caption{Computed mass (in GeV) of ground-state light baryons. $M_{exp}$ and
$\protect\mu_{exp}$ are the observed values of mass and magnetic moments
(all in $\mu_{N}$), respectively.}
\label{tab:lightbaryon}%
\begin{tabular}{ccccccc}
\hline\hline
\textrm{State} & $R_{0}$(GeV$^{-1}$) & $M_{bag}$ & $M_{exp}$ \cite{Tanabashi:D18}
& $\mu_{bag}$ & $\mu_{exp}$ \cite{Tanabashi:D18} & $r_{E}$%
(fm) \\ \hline
$p$ & 5.22 & 0.932 & 0.938 & 2.79 & 2.79 & 0.75 \\
$n$ &  &  & 0.940 & -1.86 & -1.91 & 0 \\
$\Delta^{++}$ & 5.33 & 1.241 & 1.231 & 5.70 & 6.14 & 1.08 \\
$\Delta^{+}$ &  &  & 1.235 & 2.85 & 2.7 & 0.77 \\
$\Delta^{0}$ &  &  & 1.233 & 0 & - & 0 \\
$\Delta^{-}$ &  &  & - & -2.85 & - & 0.77 \\
$\Lambda$ & 5.26 & 1.096 & 1.116 & -0.71 & -0.61 & 0.17 \\
$\Sigma^{+}$ & 5.22 & 1.137 & 1.189 & 2.72 & 2.46 & 0.77 \\
$\Sigma^{0}$ &  &  & 1.193 & 0.86 & - & 0.17 \\
$\Sigma^{-}$ &  &  & 1.197 & -1.01 & -1.16 & 0.73 \\
$\Sigma^{\ast +}$ & 5.38 & 1.383 & 1.383 & 3.11 & - & 0.79 \\
$\Sigma^{\ast 0}$ &  &  & 1.384 & 0.23 & - & 0.18 \\
$\Sigma^{\ast -}$ &  &  & 1.387 & -2.64 & - & 0.75 \\
$\Xi^{0}$ & 5.27 & 1.282 & 1.315 & -1.58 & -1.25 & 0.25 \\
$\Xi^{-}$ &  &  & 1.322 & -0.64 & -0.65 & 0.72 \\
$\Xi^{\ast 0}$ & 5.42 & 1.529 & 1.532 & 0.48 & - & 0.26 \\
$\Xi^{\ast -}$ &  &  & 1.535 & -2.43 & - & 0.74 \\
$\Omega$ & 5.46 & 1.677 & 1.672 & -2.20 & -2.02 & 0.72 \\ \hline\hline
\end{tabular}%
\end{table}

Given these values for the $QQ^{\prime }$ binding energies and the
expressions for the CMI matrices in Eq. (\ref{CMI}) with the coefficients
derived in Appendices A and B, one can numerically solve the MIT model (\ref%
{MBm}) via the variational method for all established hadrons in their
lowest-lying states and thereby compute masses, magnetic moments and charge
radii for them. The results are listed in Tables \ref{tab:lightmeson}, \ref%
{tab:heavymeson}, \ref{tab:lightbaryon}, \ref{tab:1heavybaryon} and Table %
\ref{tab:2heavybaryon} for the states without state mixing due to the CMI.
The results for the DH baryons are also presented in Tables \ref%
{tab:2heavybaryon}.

Some remarks are in order: (i) Some of the computed masses $M_{bag}$ in
Table \ref{tab:lighthadron} deviate from the measured masses about $30\sim
40$ MeV; (ii) The anti-particles of mesons are not listed in Table \ref%
{tab:lightmeson} and \ref{tab:heavymeson} as they share the same masses,
charge radii but minus magnetic moments in comparison with the mesons in
Tables. The anti-particles of the heavy tetraquarks are ignored as
well; (iii) In Table \ref{tab:2heavybaryon}, our mass predictions $3.714$
GeV(for $\Xi _{cc}^{\ast }$) are comparable to the quark-model prediction $%
M\left( \Xi _{cc}^{\ast }\right) =3.727$ GeV \cite{Ebert:2004ck}, and also
to $3706\pm 28\,\mathrm{MeV}$ and $3692\pm 28\,\mathrm{MeV}$ by the lattice
QCD \cite{PACS-CS:2013vie,Brown:2014ena} respectively. Table X shows comparison of our predictions with other works for DH baryons. The prediction $3.604
$ GeV for the $\Xi _{cc}$ is in consistent with the measured mass $3.621$
GeV, considering the simplicity of the model. (iv) The predicted magnetic
moments in Table \ref{tab:lightbaryon} are in good agreement with the
measured values, from which the magnetic moments for heavy baryons and
tetraquarks are predicted; (v) Our prediction $0.75$ fm for the proton
charge radius is slightly lower than the newly-measured value $0.83$ fm\cite%
{Bezginov:2019mdi,Xiong:2019umf}, similar to original MIT bag model \cite%
{Johnson:1975zp,DeGrand:1975cf}.

\renewcommand{\tabcolsep}{0.28cm} \renewcommand{\arraystretch}{1.1}
\begin{table}[!htb]
\caption{Computed mass(in GeV), magnetic moments(all in $\protect\mu_{N}$) and charge radii of ground-state SH baryons (non-mixed). $M_{exp}$ stands for the
observed mass isospin-averaged \protect\cite{Tanabashi:D18}. }
\label{tab:1heavybaryon}%
\begin{tabular}{ccccccc}
\hline\hline
\textrm{State} & $R_{0}$(GeV$^{-1}$) & $M_{bag}$ & $M_{exp}$ \cite{Tanabashi:D18}
& $\mu_{bag}$ & $\mu$\cite{Faessler:2006ft} & $r_{E}$(fm) \\ \hline
$\Lambda_{c}$ & 4.86 & 2.270 & 2.286 & 0.49 & 0.42 & 0.60 \\
$\Sigma_{c}^{++}$ & 4.82 & 2.411 & 2.454 & 2.13 & 1.76 & 0.92 \\
$\Sigma_{c}^{+}$ &  &  & 2.453 & 0.41 & 0.36 & 0.60 \\
$\Sigma_{c}^{0}$ &  &  & 2.454 & -1.31 & -1.04 & 0.35 \\
$\Sigma_{c}^{\ast ++}$ & 5.01 & 2.512 & 2.518 & 4.07 & - & 0.95 \\
$\Sigma_{c}^{\ast +}$ &  &  & 2.518 & 1.39 & - & 0.62 \\
$\Sigma_{c}^{\ast 0}$ &  &  & 2.518 & -1.29 & - & 0.37 \\
$\Omega_{c}$ & 4.93 & 2.680 & 2.695 & -1.07 & -0.85 & 0.28 \\
$\Omega_{c}^{\ast}$ & 5.10 & 2.764 & 2.766 & -0.90 & - & 0.29 \\
$\Lambda_{b}$ & 4.60 & 5.648 & 5.620 & -0.09 & -0.06 & 0.25 \\
$\Sigma_{b}^{+}$ & 4.64 & 5.835 & 5.811 & 2.23 & 2.07 & 0.71 \\
$\Sigma_{b}^{0}$ &  &  & - & 0.58 & 0.53 & 0.26 \\
$\Sigma_{b}^{-}$ &  &  & 5.816 & -1.07 & -1.01 & 0.62 \\
$\Sigma_{b}^{\ast +}$ & 4.73 & 5.872 & 5.830 & 3.29 & - & 0.73 \\
$\Sigma_{b}^{\ast 0}$ &  &  & - & 0.76 & - & 0.26 \\
$\Sigma_{b}^{\ast -}$ &  &  & 5.835 & -1.77 & - & 0.63 \\
$\Omega_{b}$ & 4.77 & 6.080 & 6.046 & -0.86 & -0.82 & 0.60 \\
$\Omega_{b}^{\ast}$ & 4.84 & 6.112 & - & -1.43 & - & 0.60 \\ \hline\hline
\end{tabular}%
\end{table}

\section{Baryons and Tetraquarks}

\label{Hadrons}

\subsection{Heavy Baryons including the CMI Mixing}

Hadrons containing a diquark or antiquark with different flavors, may not
respect flavor-symmetry of wavefunction for involved light quark pairs. As
such, the states with same $J^{P}$ but different spin-color wavefunctions
may mix due to the CMI (\ref{CMI}), as mentioned in Sect. II (B). To begin
with, we first consider the system of baryons with $J^{P}={1/2}^{+}$ in
which two spin-color states $(\phi ^{B}\chi _{2}^{B},\phi ^{B}\chi _{3}^{B})$ can
mix. The associated baryons are the $\Xi _{c}$, the $\Xi _{b}$, the $\Xi
_{bc}$ and the $\Omega _{bc}$.

\renewcommand{\tabcolsep}{0.46cm} \renewcommand{\arraystretch}{1.1}
\begin{table*}[!htb]
\caption{Predicted masses (in GeV), magnetic moments(in $\protect\mu_{N}$) and charge radii of 
heavy baryons. $M_{exp}$ is the observed mass isospin-averaged \protect\cite%
{Tanabashi:D18}. Magnetic moments and charge radii are organized in the
order of $I_{3}=\frac{1}{2},\,-\frac{1}{2}$ for $I=\frac{1}{2}$. Bag radius $%
R_{0}$ is in GeV$^{-1}$.}
\label{tab:mixingbaryon}%
\begin{tabular}{ccccccc}
\hline\hline
\textrm{State} & \textrm{Eigenvector} & $R_{0}$ & $M_{bag}$ & $M_{exp}$ \cite%
{Tanabashi:D18} & $\mu_{bag}$ & $r_{E}$(fm) \\ \hline
\multirow{2}{*}{$\left(\Xi_{c}^{\prime},\ \Xi_{c}\right)$} & (0.05, 1.00) &
4.89 & 2.436 & 2.469 & 0.37, 0.50 & 0.63, 0.32 \\
& (-1.00, 0.05) & 4.88 & 2.544 & 2.578 & 0.67, -1.20 & 0.63, 0.32 \\
$\Xi_{c}^{\ast}$ & 1.00 & 5.06 & 2.636 & 2.646 & 1.61, -1.10 & 0.65, 0.33 \\
\multirow{2}{*}{$\left(\Xi_{b}^{\prime},\ \Xi_{b}\right)$} & (0.01, 1.00) &
4.64 & 5.805 & 5.794 & -0.12, -0.08 & 0.30, 0.60 \\
& (-1.00, 0.01) & 4.71 & 5.956 & 5.935 & 0.74, -0.97 & 0.30, 0.61 \\
$\Xi_{b}^{\ast}$ & 1.00 & 4.79 & 5.991 & 5.954 & 0.96, -1.61 & 0.31, 0.62 \\
\multirow{2}{*}{$\left(\Xi_{bc},\ \Xi_{bc}^{\prime}\right)$} & (0.39, 0.92)
& 4.22 & 7.015 & - & 1.48, -0.33 & 0.58, 0.19 \\
& (-0.92, 0.39) & 4.09 & 6.953 & - & -0.20, 0.09 & 0.56, 0.19 \\
$\Xi_{bc}^{\ast}$ & 1.000 & 4.31 & 7.044 & - & 1.94, -0.37 & 0.59, 0.20 \\
\multirow{2}{*}{$\left(\Omega_{bc},\ \Omega_{bc}^{\prime}\right)$} & (0.40,
0.92) & 4.29 & 7.117 & - & -0.20 & 0.15 \\
& (-0.92, 0.40) & 4.18 & 7.064 & - & 0.06 & 0.14 \\
$\Omega_{bc}^{\ast}$ & 1.000 & 4.37 & 7.143 & - & -0.22 & 0.15 \\
\hline\hline
\end{tabular}%
\end{table*}

\renewcommand{\tabcolsep}{0.46cm} \renewcommand{\arraystretch}{1.1}
\begin{table}[!htb]
\setlength{\abovecaptionskip}{0.5cm}
\caption{Computed masses (in GeV), magnetic moments(all in $\protect\mu_{N}$) and charge radii
of the ground-state doubly heavy baryons (non-mixed states).
The magnetic moments by Ref.~\cite{Faessler:2006ft} are listed for comparison.}
\label{tab:2heavybaryon}%
\begin{tabular}{cccccc}
\hline\hline
\textrm{State} & $R_{0}$(GeV$^{-1}$) & $M_{bag}$ & $\mu_{bag}$
& $\mu$\cite{Faessler:2006ft} & $r_{E}$(fm) \\ \hline
$\Xi_{cc}^{++}$ & 4.42 & 3.604 & 0.12 & 0.13 & 0.78 \\
$\Xi_{cc}^{+}$ &  &  & 0.91 & 0.72 & 0.45 \\
$\Xi_{cc}^{\ast ++}$ & 4.64 & 3.714 & 2.64 & - & 0.82 \\
$\Xi_{cc}^{\ast +}$ &  &  & 0.16 & - & 0.47 \\
$\Xi_{bb}^{0}$ & 3.71 & 10.311 & -0.55 & -0.53 & 0.29 \\
$\Xi_{bb}^{-}$ &  &  & 0.11 & 0.18 & 0.45 \\
$\Xi_{bb}^{\ast 0}$ & 3.87 & 10.360 & 1.21 & - & 0.30 \\
$\Xi_{bb}^{\ast -}$ &  &  & -0.86 & - & 0.47 \\
$\Omega_{cc}$ & 4.49 & 3.726 & 0.86 & 0.67 & 0.48 \\
$\Omega_{cc}^{\ast}$ & 4.69 & 3.820 & 0.33 & - & 0.50 \\
$\Omega_{bb}$ & 3.83 & 10.408 & 0.07 & 0.04 & 0.45 \\
$\Omega_{bb}^{\ast}$ & 3.97 & 10.451 & -0.75 & - & 0.47 \\ \hline\hline
\end{tabular}%
\end{table}

\renewcommand{\tabcolsep}{0.34cm} \renewcommand{\arraystretch}{1.1}
\begin{table*}[!hbt]
\caption{Computed mass and other calculations cited of doubly heavy baryons, all in GeV.}
\label{tab:comparisonB}%
\begin{tabular}{cccccccccc}
\hline\hline
\textrm{State} & $J$ & \textrm{This\ work} & \cite{KR:2014gca} & \cite%
{Aliev:2012iv} & \cite{Ebert:2004ck} & \cite{Roberts:2007ni} & \cite{Albertus:2006ya} &
\cite{Giannuzzi:2009gh} & \cite{Bernotas:2008fv} \\ \hline
$\Xi_{cc}$ & $\frac{1}{2}$ & 3.604 & 3.627 & - & 3.620 & 3.676 & 3.612 & 3.547 &
3.557 \\
$\Xi_{cc}^{\ast}$ & $\frac{3}{2}$ & 3.714 & 3.690 & 3.72 & 3.727 & 3.753 & 3.706 &
3.719 & 3.661 \\
$\Omega_{cc}$ & $\frac{1}{2}$ & 3.726 & - & - & 3.778 & 3.815 & 3.702 & 3.648 &
3.710 \\
$\Omega_{cc}^{\ast}$ & $\frac{3}{2}$ & 3.820 & - & 3.78 & 3.872 & 3.876 & 3.783 &
3.770 & 3.800 \\
$\Xi_{bb}$ & $\frac{1}{2}$ & 10.311 & 10.162 & - & 10.202 & 10.340 & 10.197 &
10.185 & 10.062 \\
$\Xi_{bb}^{\ast}$ & $\frac{3}{2}$ & 10.360 & 10.184 & 10.3 & 10.237 & 10.367 &
10.236 & 10.216 & 10.101 \\
$\Omega_{bb}$ & $\frac{1}{2}$ & 10.408 & - & - & 10.359 & 10.454 & 10.260 &
10.271 & 10.208 \\
$\Omega_{bb}^{\ast}$ & $\frac{3}{2}$ & 10.451 & - & 10.4 & 10.389 & 10.486 & 10.297
& 10.289 & 10.244 \\
$\Xi_{bc}$ & $\frac{1}{2}$ & 6.953 & 6.914 & - & 6.933 & 7.011 & 6.919 & 6.904 &
6.846 \\
$\Xi_{bc}^{\prime}$ & $\frac{1}{2}$ & 7.015 & 6.933 & - & 6.963 & 7.047 & 6.948
& 6.920 & 6.891 \\
$\Xi_{bc}^{\ast}$ & $\frac{3}{2}$ & 7.044 & 6.969 & 7.2 & 6.980 & 7.074 & 6.986 &
6.936 & 6.919 \\
$\Omega_{bc}$ & $\frac{1}{2}$ & 7.064 & - & - & 7.088 & 7.136 & 6.986 & 6.994 &
6.999 \\
$\Omega_{bc}^{\prime}$ & $\frac{1}{2}$ & 7.116 & - & - & 7.116 & 7.165 & 7.009 &
7.005 & 7.036 \\
$\Omega_{bc}^{\ast}$ & $\frac{3}{2}$ & 7.142 & - & 7.35 & 7.130 & 7.187 & 7.046 &
7.017 & 7.063 \\ \hline\hline
\end{tabular}%
\end{table*}

\renewcommand{\tabcolsep}{0.37cm} \renewcommand{\arraystretch}{1.1}
\begin{table*}[!htb]
\caption{Computed mass (in GeV), magnetic moments(in $\protect\mu_{N}$) and charge radii of singly heavy tetraquarks $nn\bar{s}\bar{c}$ and $nn\bar{s}\bar{b}$. Magnetic
moments and charge radii are organized in the order of $I_{3}=1,\,0,\,-1$
for $I=1$. Bag radius $R_{0}$ is in GeV$^{-1}$. }
\label{tab:1heavytetraquark}%
\begin{tabular}{ccccccc}
\hline\hline
\textrm{State} & $J^{P}$ & \textrm{Eigenvector} & $R_{0}$ & $M_{bag}$ & $%
\mu_{bag}$ & $r_{E}$ (fm) \\ \hline
\multirow{2}{*}{${(nn\bar{s}\bar{c})}^{I=1}$} & \multirow{2}{*}{$0^{+}$} &
(0.54, 0.84) & 5.73 & 3.218 & - & 0.91, 0.38, 0.73 \\
&  & (-0.84, 0.55) & 5.39 & 2.776 & - & 0.85, 0.35, 0.69 \\
& \multirow{3}{*}{$1^{+}$} & (0.81, 0.58, 0.10) & 5.46 & 3.001 & 3.48, 1.54,
-0.40 & 0.86, 0.36, 0.70 \\
&  & (0.25, -0.49, 0.84) & 5.57 & 3.154 & 1.03, 0.23, -0.57 & 0.88, 0.37,
0.71 \\
&  & (-0.54, 0.65, 0.54) & 5.38 & 2.846 & 1.67, 0.04, -1.60 & 0.85, 0.35,
0.69 \\
& $2^{+}$ & 1.00 & 5.64 & 3.075 & 4.27, 1.25, -1.77 & 0.89, 0.37, 0.72 \\
\multirow{2}{*}{${(nn\bar{s}\bar{c})}^{I=0}$} & \multirow{2}{*}{$0^{+}$} &
(0.63, 0.77) & 5.56 & 2.934 & - & 0.37 \\
&  & (-0.78, 0.63) & 5.19 & 2.513 & - & 0.34 \\
& \multirow{3}{*}{$1^{+}$} & (0.77, 0.07, 0.64) & 5.40 & 2.895 & 0.54 & 0.35
\\
&  & (-0.23, -0.90, 0.36) & 5.57 & 3.056 & 1.24 & 0.37 \\
&  & (-0.60, 0.43, 0.68) & 5.35 & 2.674 & 0.04 & 0.35 \\
& $2^{+}$ & 1.00 & 5.66 & 3.063 & 1.26 & 0.37 \\
\multirow{2}{*}{${(nn\bar{s}\bar{b})}^{I=1}$} & \multirow{2}{*}{$0^{+}$} &
(0.53, 0.85) & 5.53 & 6.580 & - & 1.07, 0.71, 0.36 \\
&  & (-0.85, 0.53) & 5.28 & 6.202 & - & 1.02, 0.68, 0.34 \\
& \multirow{3}{*}{$1^{+}$} & (0.66, 0.75, 0.06) & 5.35 & 6.419 & 3.59, 1.37,
-0.86 & 1.03, 0.69, 0.35 \\
&  & (0.36, -0.38, 0.85) & 5.43 & 6.554 & 1.33, 0.72, 0.12 & 1.05, 0.70, 0.35
\\
&  & (-0.66, 0.55, 0.52) & 5.22 & 6.228 & 1.99, 0.55, -0.89 & 1.01, 0.67,
0.34 \\
& $2^{+}$ & 1.00 & 5.47 & 6.446 & 4.72, 1.79, -1.13 & 1.05, 0.70, 0.36 \\
\multirow{2}{*}{${(nn\bar{s}\bar{b})}^{I=0}$} & \multirow{2}{*}{$0^{+}$} &
(0.66, 0.75) & 5.41 & 6.327 & - & 0.69 \\
&  & (-0.75, 0.66) & 5.14 & 5.980 & - & 0.66 \\
& \multirow{3}{*}{$1^{+}$} & (0.66, -0.23, 0.72) & 5.33 & 6.322 & 0.71 & 0.68
\\
&  & (-0.46, -0.88, 0.14) & 5.38 & 6.454 & 1.31 & 0.69 \\
&  & (-0.60, 0.43, 0.68) & 5.14 & 6.038 & 0.61 & 0.66 \\
& $2^{+}$ & 1.00 & 5.48 & 6.431 & 1.80 & 0.70 \\ \hline\hline
\end{tabular}%
\end{table*}

In terms of the wavefunctions in color and spin space (Appendix A and B),
one can compute the CMI matrices in the degenerate subspace of the
spin-color basis $\phi ^{B}\chi ^{B}$ when the chromomagnetic mixing occurs
(Appendix C). These CMI matrices depend upon $C_{ij}$ with the subscripts $%
(i,j)$ of $C_{ij}$ denote the flavor constituents. One can diagonalize the
CMI matrix, say (\ref{mixing23}), to write mass formulas of the baryons
using Eq. (\ref{MBm}). This is done by solving the eigenvalues and
eigenvectors of the matrix (\ref{mixing23}) analytically and using the later
to identify(denote) the mixed states. Of course, the relevant binding
energies ($B_{cs}$, $B_{bs}$ and $B_{bc}$) are included in the mass formulas.

In Table \ref{tab:mixingbaryon}, we list our computed results of masses and
other properties for the CMI-mixed systems of heavy baryons. The net effects
of the state mixing (the second column of Table) are not so significant in
general and they are somehow negligible in the case of singly heavy baryons.
This can be due to the higher $SU(3)$ flavor symmetry and heavy quark
symmetry which suppress the off-diagonal elements in matrix (\ref{mixing23}%
). For this reason, we employ still the normal notations of the states for
the SH baryons. The computed masses of the SH baryons $\Xi _{c}$, the $\Xi
_{c}^{\prime }$, the $\Xi _{b}$ and the $\Xi _{b}^{\prime }$ are comparable
with the measured data, as seen in the fifth column with reasonable errors.
The magnetic moments for $\Xi_{c}^{+}$, $\Xi_{c}^{0}$, $\Xi_{b}^{0}$ and
$\Xi_{b}^{-}$ are predicted to be $0.37\mu_{N}$, $0.50\mu_{N}$, $-0.12\mu_{N}$
and $-0.08\mu_{N}$ which are comparable to $0.35\mu_{N}$, $0.50\mu_{N}$,
$-0.045\mu_{N}$ and $-0.08\mu_{N}$ in Ref.~\cite{Aliev:2008ay}, respectively.
The magnetic moments for $\Xi_{c}^{\ast +}$, $\Xi_{c}^{\ast 0}$,
$\Xi_{b}^{\ast 0}$ and $\Xi_{b}^{\ast -}$ are $1.61\mu_{N}$, $-1.10\mu_{N}$,
$0.96\mu_{N}$ and $-1.61\mu_{N}$ comparable to $1.68\mu_{N}$, $-0.68\mu_{N}$,
$0.50\mu_{N}$ and $-1.42\mu_{N}$ in Ref.~\cite{Aliev:2008sk}, respectively.

\subsection{Singly Heavy Tetraquarks}

Let us consider the strange tetraquarks $nn\bar{s}\bar{c}$ and $nn\bar{s}%
\bar{b}$ containing one heavy quark, one strange quark and two nonstrange
light quarks. In such a case, the CMI mixing happens if $J\neq 2$. We use a
combination of the spin-color basis functions $\phi ^{T}\chi ^{T}$ to denote
the mixed states. For instance, the combination $(\phi _{2}^{T}\chi
_{3}^{T},\phi _{1}^{T}\chi _{6}^{T})$ stands for a mixed state $c_{1}\phi
_{2}^{T}\chi _{3}^{T}+c_{2}\phi _{1}^{T}\chi _{6}^{T}$ for ($J^{P},I$)$=$($%
0^{+}$ , $1$). Similarly, other mixed states can be denoted as $(\phi
_{1}^{T}\chi _{3}^{T},\phi _{2}^{T}\chi _{6}^{T})$ for ($J^{P},I$)$=$($0^{+}$
, $0$), as $(\phi _{2}^{T}\chi _{2}^{T},\phi _{2}^{T}\chi _{4}^{T},\phi
_{1}^{T}\chi _{5}^{T})$ for ($J^{P},I$)$=$($1^{+}$ , $1$) and $(\phi
_{1}^{T}\chi _{2}^{T},\phi _{1}^{T}\chi _{4}^{T},\phi _{2}^{T}\chi _{5}^{T})$
for ($J^{P},I$)$=$($1^{+}$ , $0$). The binding energy matrices become
diagonal in mass formula since the mixed states have two color
configurations while spin states are orthogonal. Note that diagonalization
should be applied to the sum of the interaction matrices before evaluating
the hadron mass.

Following the variational principle, we diagonalize the $2\times 2$ matrix
to solve two analytical eigenvalues and construct the mass formula as usual.
Application of the same procedure to the $3\times 3$ matrix is, however, not
straightforward, for which the eigenvalues are some roots of a cubic
equation. For this, we scan three sets of $x_{i}$ and $R$ to solve the cubic
equation numerically so that one can obtain the minimized masses within
three root eigenvalues.

\renewcommand{\tabcolsep}{0.33cm} \renewcommand{\arraystretch}{1.1}
\begin{table*}[!htb]
\caption{Computed mass (in GeV) and other properties of doubly heavy
tetraquarks $nn\bar{c}\bar{c}$, $nn\bar{b}\bar{b}$ and $nn\bar{c}\bar{b}$.
Magnetic moments(in $\protect\mu_{N}$) and charge radii are organized in the order of $I_{3}=1,\,0,\,-1$ for $I=1$. Bag radius $R_{0}$ is in GeV$^{-1}$.
The mass and magnetic moment of $T_{cc}^{+}$ are predicted to be
$3.925\,{\rm GeV}$ and $0.88\mu_{N}$ which is comparable to
$0.66\mu_{N}$ in Ref. \cite{Azizi:2021aib}. }
\label{tab:2heavytetraquark1}%
\begin{tabular}{ccccccc}
\hline\hline
\textrm{State} & $J^{P}$ & \textrm{Eigenvector} & $R_{0}$ & $M_{bag}$ & $%
\mu_{bag}$ & $r_{E}$ (fm) \\ \hline
\multirow{2}{*}{${(nn\bar{c}\bar{c})}^{I=1}$} & \multirow{2}{*}{$0^{+}$} &
(0.40, 0.92) & 5.40 & 4.342 & - & 0.56, 0.54, 0.94 \\
&  & (-0.91, 0.41) & 5.04 & 4.032 & - & 0.52, 0.50, 0.88 \\
& $1^{+}$ & 1.00 & 5.22 & 4.117 & 1.36, -0.03, -1.43 & 0.54, 0.52, 0.91 \\
& $2^{+}$ & 1.00 & 5.32 & 4.179 & 2.80, -0.05, -2.90 & 0.55, 0.53, 0.93 \\
\multirow{2}{*}{${(nn\bar{c}\bar{c})}^{I=0}$} & \multirow{2}{*}{$1^{+}$} &
(0.97, 0.25) & 5.15 & 3.925 & -0.88 & 0.51 \\
&  & (-0.24, 0.97) & 5.30 & 4.205 & 0.83 & 0.53 \\
\multirow{2}{*}{${(nn\bar{b}\bar{b})}^{I=1}$} & \multirow{2}{*}{$0^{+}$} &
(0.17, 0.99) & 4.90 & 11.092 & - & 0.92, 0.59, 0.38 \\
&  & (-0.98, 0.18) & 4.77 & 10.834 & - & 0.90, 0.58, 0.37 \\
& $1^{+}$ & 1.00 & 4.83 & 10.854 & 1.81, 0.52, -0.78 & 0.91, 0.58, 0.38 \\
& $2^{+}$ & 1.00 & 4.88 & 10.878 & 3.65, 1.04, -1.57 & 0.92, 0.59, 0.38 \\
\multirow{2}{*}{${(nn\bar{b}\bar{b})}^{I=0}$} & \multirow{2}{*}{$1^{+}$} &
(1.00, 0.08) & 4.76 & 10.654 & 0.18 & 0.57 \\
&  & (-0.08, 1.00) & 4.83 & 10.982 & 0.86 & 0.58 \\
\multirow{2}{*}{${(nn\bar{c}\bar{b})}^{I=1}$} & \multirow{2}{*}{$0^{+}$} &
(0.30, 0.95) & 5.16 & 7.714 & - & 0.78, 0.25, 0.70 \\
&  & (-0.95, 0.31) & 4.91 & 7.438 & - & 0.74, 0.23, 0.67 \\
& \multirow{3}{*}{$1^{+}$} & (0.65, 0.76, 0.09) & 5.04 & 7.509 & 2.33, 0.21,
-1.90 & 0.76, 0.24, 0.68 \\
&  & (0.13, -0.22, 0.97) & 5.10 & 7.699 & -0.17, -0.32, -0.47 & 0.77, 0.24,
0.69 \\
&  & (-0.75, 0.61, 0.24) & 4.96 & 7.465 & 2.57, 0.82, -0.93 & 0.75, 0.24,
0.67 \\
& $2^{+}$ & 1.00 & 5.12 & 7.531 & 3.24, 0.50, -2.24 & 0.78, 0.25, 0.69 \\
\multirow{2}{*}{${(nn\bar{c}\bar{b})}^{I=0}$} & \multirow{2}{*}{$0^{+}$} &
(0.93, 0.37) & 4.96 & 7.502 & - & 0.24 \\
&  & (-0.38, 0.93) & 4.84 & 7.260 & - & 0.23 \\
& \multirow{3}{*}{$1^{+}$} & (0.91, -0.36, 0.21) & 4.93 & 7.518 & 0.55 & 0.23
\\
&  & (-0.39, -0.92, 0.10) & 5.07 & 7.605 & 0.50 & 0.24 \\
&  & (-0.16, 0.18, 0.97) & 4.93 & 7.288 & -0.33 & 0.23 \\
& $2^{+}$ & 1.00 & 5.14 & 7.483 & 0.50 & 0.25 \\ \hline\hline
\end{tabular}%
\end{table*}

Our numerical results are shown in Table \ref{tab:1heavytetraquark}, with a
notable tetraquark of an isosinglet $nn\bar{s}\bar{c}$ with $J^{P}=0^{+}$,
which has two masses $2.934$ GeV and $2.513$ GeV for its two mixed states.
Comparing with the measured mass $2866\pm 7$ MeV of $X_{0}(2900)$ reported
by LHCb\cite{LHCb:2020kd} and the quark model prediction $2863.4\pm 12$ MeV%
\cite{Karliner:2020vsi}, our prediction $2.934$ GeV is larger even if the
model error $40$ MeV is subtracted. If we rather, as Karliner suggested for
the color $\boldsymbol{\bar{3}}\otimes \boldsymbol{3}$ configuration, ignore
the CMI mixing and evaluate directly the masses of the $\phi _{1}^{T}\chi
_{3}^{T}$ and $\phi _{2}^{T}\chi _{6}^{T}$ states, the resulted masses lie
around $2.7$ GeV, away from the LHCb reported mass of the $X_{0}(2900)$. Our
calculation suggests that chromomagnetic mixing is strong for the strange
tetraquark $nn\bar{s}\bar{c}$ with $J^{P}=0^{+}$ and yields a mass splitting
as large as $420$ MeV.

\renewcommand{\tabcolsep}{0.60cm} \renewcommand{\arraystretch}{1.1}
\begin{table*}[!htb]
\caption{Computed mass (in GeV), magnetic moments(in $\protect\mu_{N}$) and charge radii of doubly heavy tetraquarks $ss\bar{c}\bar{c}$, $ss\bar{b}\bar{b}$ and $ss\bar{c}\bar{b}$. Bag radius $R_{0}$ is in GeV$^{-1}$. }
\label{tab:2heavytetraquark0}%
\begin{tabular}{ccccccc}
\hline\hline
\textrm{State} & $J^{P}$ & \textrm{Eigenvector} & $R_{0}$ & $M_{bag}$ & $%
\mu_{bag}$ & $r_{E}$ (fm) \\ \hline
\multirow{2}{*}{$ss\bar{c}\bar{c}$} & \multirow{2}{*}{$0^{+}$} & (0.49, 0.87)
& 5.48 & 4.521 & - & 0.92 \\
&  & (-0.87, 0.50) & 5.13 & 4.300 & - & 0.87 \\
& $1^{+}$ & 1.00 & 5.30 & 4.382 & -1.22 & 0.89 \\
& $2^{+}$ & 1.00 & 5.39 & 4.433 & -2.46 & 0.91 \\
\multirow{2}{*}{$ss\bar{b}\bar{b}$} & \multirow{2}{*}{$0^{+}$} & (0.24, 0.97)
& 5.01 & 11.232 & - & 0.32 \\
&  & (-0.97, 0.25) & 4.88 & 11.078 & - & 0.31 \\
& $1^{+}$ & 1.00 & 4.94 & 11.099 & -0.60 & 0.31 \\
& $2^{+}$ & 1.00 & 4.98 & 11.119 & -1.20 & 0.32 \\
\multirow{2}{*}{$ss\bar{c}\bar{b}$} & \multirow{2}{*}{$0^{+}$} & (0.40, 0.92)
& 5.26 & 7.875 & - & 0.67 \\
&  & (-0.91, 0.40) & 5.01 & 7.693 & - & 0.64 \\
& \multirow{3}{*}{$1^{+}$} & (0.70, 0.71, 0.11) & 4.98 & 7.757 & -1.54 & 0.64
\\
&  & (0.17, -0.32, 0.93) & 5.06 & 7.858 & -0.48 & 0.64 \\
&  & (-0.69, 0.63, 0.35) & 4.88 & 7.716 & -0.66 & 0.62 \\
& $2^{+}$ & 1.00 & 5.20 & 7.779 & -1.84 & 0.66 \\ \hline\hline
\end{tabular}%
\end{table*}

\renewcommand{\tabcolsep}{0.52cm} \renewcommand{\arraystretch}{1.1}
\begin{table*}[!htb]
\setlength{\abovecaptionskip}{0.4cm}
\caption{Computed mass (in GeV) and other properties of doubly heavy
tetraquarks $ns\bar{c}\bar{c}$, $ns\bar{b}\bar{b}$. Magnetic moments(in $\protect\mu_{N}$) and charge radii are organized in the order of $I_{3}=\frac{1}{2},\,-\frac{1}{2}$
for $I=\frac{1}{2}$. Bag radius $R_{0}$ is in GeV$^{-1}$. }
\label{tab:2heavytetraquark1/2}%
\begin{tabular}{ccccccc}
\hline\hline
\textrm{State} & $J^{P}$ & \textrm{Eigenvector} & $R_{0}$ & $M_{bag}$ & $%
\mu_{bag}$ & $r_{E}$ (fm) \\ \hline
\multirow{2}{*}{$ns\bar{c}\bar{c}$} & \multirow{2}{*}{$0^{+}$} & (0.44, 0.90)
& 5.44 & 4.429 & - & 0.51, 0.93 \\
&  & (-0.89, 0.45) & 5.09 & 4.165 & - & 0.48, 0.88 \\
& \multirow{3}{*}{$1^{+}$} & (0.99, -0.07, 0.09) & 5.16 & 4.247 & 0.32, -1.33
& 0.49, 0.89 \\
&  & (-0.11, -0.28, 0.95) & 5.23 & 4.314 & 0.86, -1.58 & 0.49, 0.90 \\
&  & (0.04, 0.96, 0.29) & 5.04 & 4.091 & -0.95, -1.03 & 0.48, 0.87 \\
& $2^{+}$ & 1.00 & 5.36 & 4.305 & 0.19, -2.68 & 0.50, 0.92 \\
\multirow{2}{*}{$ns\bar{b}\bar{b}$} & \multirow{2}{*}{$0^{+}$} & (0.20, 0.98)
& 4.96 & 11.160 & - & 0.62, 0.35 \\
&  & (-0.98, 0.20) & 4.83 & 10.955 & - & 0.60, 0.34 \\
& \multirow{3}{*}{$1^{+}$} & (1.00, -0.01, 0.03) & 4.79 & 10.974 & 0.65,
-0.68 & 0.60, 0.34 \\
&  & (-0.03, -0.09, 1.00) & 4.77 & 11.068 & 1.02, -1.50 & 0.60, 0.34 \\
&  & (0.01, 1.00, 0.10) & 4.66 & 10.811 & 0.14, 0.16 & 0.58, 0.33 \\
& $2^{+}$ & 1.00 & 4.93 & 10.997 & 1.25, -1.39 & 0.62, 0.35 \\ \hline\hline
\end{tabular}%
\end{table*}

\subsection{Doubly Heavy Tetraquarks}

Now, let us consider the doubly heavy tetraquarks $qq\bar{Q}\bar{Q}$ with
strangeness $S\leq 2$. In this case, hadrons consist of the nonstrange
tetraquarks $nn\bar{Q}\bar{Q}$ and the strange tetraquarks $ns\bar{Q}\bar{Q}$
and $ss\bar{Q}\bar{Q}$. They lie in a larger(compared to baryons) space
spanned by more configurations(bases) in which the CMI mixing occurs
variously. For the isotriplet tetraquarks with $J^{P}=0^{+}$, the general
ground state can be the mixed one, with the wavefunction $(\phi _{2}^{T}\chi
_{3}^{T},\phi _{1}^{T}\chi _{6}^{T})$ . For isosinglet tetraquark $nn\bar{Q}%
\bar{Q}$ with $J^{P}=1^{+}$, the wavefunction has the form of $(\phi
_{2}^{T}\chi _{5}^{T},\phi _{1}^{T}\chi _{4}^{T})$. In the case of the
strange tetraquark $ns\bar{Q}\bar{Q}$ with $J^{P}=1^{+}$, the wavefunction
can be of $(\phi _{2}^{T}\chi _{2}^{T},\phi _{2}^{T}\chi _{5}^{T},\phi
_{1}^{T}\chi _{4}^{T})$ and the wavefunction of the tetraquark $nn$$\bar{c}%
\bar{b}$ is similar to that of $nn\bar{s}\bar{c}$. Note that the strange DH states $ns\bar{c}\bar{b}$ with mixing among six spin-color states are not considered for simplicity.

The computation of the mass and other properties of these DH tetraquarks is
similar to that for heavy baryons discussed in Sect. IV(A). Our numerical
results for DH tetraquarks are listed in Table \ref{tab:2heavytetraquark1}
for the $nn\bar{Q}\bar{Q\prime }$, in Table \ref{tab:2heavytetraquark0} for
the $ss\bar{Q}\bar{Q\prime }$ and Table \ref{tab:2heavytetraquark1/2} for
the $ns\bar{Q}\bar{Q\prime }$. Our results for the mass predictions of the
DH tetraquarks are summarized in Table \ref{tab:comparison} and compared
with some other calculations cited.

\section{Summary and Discussions}

\label{Discussions} In this work, we have studied systematically masses and
other properties of hadrons with one and two open heavy quarks within an
unified framework of MIT bag model with chromomagnetic interaction. Masses, magnetic moments and charge radii of heavy baryons and heavy tetraquarks are computed
systematically, including the predictions $M(\Xi _{cc},1/2^{+})=3.604$ GeV, $%
M(\Xi _{cc}^{\ast },,3/2^{+})=3.714$ GeV, and $M\left( ud\bar{s}\bar{c}%
,0^{+}\right) =2.934$ GeV for the strange isosinglet tetraquark $ud\bar{s}%
\bar{c}$. The state mixing due to chromomagnetic interaction is shown to be
sizable for the strange scalar tetraquark $nn\bar{s}\bar{c}$, giving mass
splitting as large as $420$ MeV roughly, while it is small for other heavy
hadrons.

We also confirm that a term of extra binding energy $B_{QQ^{\prime }}$,
proposed previously to exist among heavy quarks ($c$ and $b$) and between
heavy and strange quarks\cite{KR:2014gca}, is required to reconcile light
hadron with heavy hadrons, with a useful formula provided for $B_{QQ^{\prime
}}$. This binding effect may rise from the enhanced short-range interaction
between two relatively heavy quarks and makes the mass pattern and other
properties of heavy hadrons differing from that in light sector. We have
also employed a slowly-running strong coupling $\alpha _{s}(R)$ to reflect
its dependence upon the hadron sizes proportional to the average distance
between two interacted quarks (or antiquark) in a hadron. The strong coupling
$\alpha _{s}(R)$ runs from $0.4$ to $0.6$ as the bag radius $R$ varies
between $3\sim 6$ GeV$^{-1}$.

We remark that the MIT bag model can reproduce the measured masses of heavy
hadrons within the accuracy of $40$-$50$ MeV, from which we proceed to
predict the masses and other properties of the tetraquarks with one and two
open heavy quarks. For the DH tetraquarks, we reduce the error limit to
about $40$ MeV and exclude $X_{0}(2900)$ to be an isosinglet tetraquark of $%
nn\bar{s}\bar{c}$ due to the mismatch with the measured data as high as $70$
MeV.

Owing to the uncertainty of model computations, we are not able to discuss
the near-threshold effect. The mismatch of our predictions with the measured
data may come from the limitations of bag model in this work: (1) the bag
may deform into elliptic shape in the case of the DH hadrons, and (2) the
constant approximation of the short-range binding energy may not be
sufficient as the later may depend upon hadrons size $R$ implicitly, for
instance, in the form of a Coulomb-like $\sim 1/R$, and needs to be
determined variationally. These effects go beyond the scope of this work and
await the further exploration in the future.

\renewcommand{\tabcolsep}{0.26cm} \renewcommand{\arraystretch}{1.1}
\begin{table*}[!htb]
\caption{Comparison of calculated mass(in GeV) among different calculations
for double heavy tetraquarks. The masses before and after slash stand for
that of the color states split chromo-magnetically. Refs.\protect\cite%
{Luo:2017eub} and \protect\cite{Lu:2020rog} employ the CMI model. }
\label{tab:comparison}%
\begin{tabular}{ccccccc}
\hline\hline
\textrm{State} & $J$ & \textrm{This\ work} & \textrm{Scheme\ 1}\cite%
{Luo:2017eub} & \textrm{Scheme\ 2}\cite{Luo:2017eub} & \cite{Lu:2020rog} &
\cite{Ebert:2007rn} \\ \hline
${(nn\bar{c}\bar{c})}^{I=1}$ & 0 & 4.032/4.342 & 4.078/4.356 & 3.850/4.128 &
4.195/4.414 & 4.056 \\
& 1 & 4.117 & 4.201 & 3.973 & 4.268 & 4.079 \\
& 2 & 4.179 & 4.271 & 4.044 & 4.318 & 4.118 \\
${(nn\bar{c}\bar{c})}^{I=0}$ & 1 & 3.925/4.205 & 4.007/4.204 & 3.779/3.977 &
4.041/4.313 & 3.935 \\
${(nn\bar{b}\bar{b})}^{I=1}$ & 0 & 10.834/11.092 & 10.841/10.937 &
10.637/10.734 & 10.765/11.019 & 10.648 \\
& 1 & 10.854 & 10.875 & 10.671 & 10.779 & 10.657 \\
& 2 & 10.878 & 10.897 & 10.694 & 10.799 & 10.673 \\
${(nn\bar{b}\bar{b})}^{I=0}$ & 1 & 10.654/10.982 & 10.686/10.821 &
10.483/10.617 & 10.550/10.951 & 10.502 \\
${(nn\bar{c}\bar{b})}^{I=1}$ & 0 & 7.438/7.714 & 7.457/7.643 & 7.241/7.428 &
7.519/7.740 & 7.383 \\
& \multirow{2}{*}{1} & 7.465/7.509 & 7.473/7.548 & 7.258/7.332 &
7.537/7.561 & \multirow{2}{*}{7.396/7.403} \\
& & 7.699 & 7.609 & 7.393 & 7.729 & \\
& 2 & 7.531 & 7.582 & 7.367 & 7.586 & 7.422 \\
${(nn\bar{c}\bar{b})}^{I=0}$ & 0 & 7.260/7.502 & 7.256/7.429 & 7.041/7.213 &
7.297/7.580 & 7.239 \\
& \multirow{2}{*}{1} & 7.288/7.518 & 7.321/7.431 & 7.106/7.215 &
7.325/7.607 & \multirow{2}{*}{7.246} \\
& & 7.605 & 7.516 & 7.301 & 7.666 & \\
& 2 & 7.483 & 7.530 & 7.315 & 7.697 & - \\
$ns\bar{c}\bar{c}$ & 0 & 4.165/4.429 & 4.236/4.514 & 3.933/4.210 &
4.323/4.512 & 4.221 \\
& \multirow{2}{*}{1} & 4.091/4.247 & 4.225/4.363 & 3.921/4.060 &
4.232/4.394 & \multirow{2}{*}{4.143/4.239} \\
& & 4.314 & 4.400 & 4.096 & 4.427 & \\
& 2 & 4.305 & 4.434 & 4.131 & 4.440 & 4.271 \\
$ns\bar{b}\bar{b}$ & 0 & 10.955/11.160 & 10.999/11.095 & 10.707/10.804 &
10.883/11.098 & 10.802 \\
& \multirow{2}{*}{1} & 10.811/10.974 & 10.911/11.010 & 10.619/10.718 &
10.734/10.897 & \multirow{2}{*}{10.706/10.809} \\
& & 11.068 & 11.037 & 10.745 & 11.046 & \\
& 2 & 10.997 & 11.060 & 10.769 & 10.915 & 10.823 \\
$ss\bar{c}\bar{c}$ & 0 & 4.300/4.521 & 4.395/4.672 & 4.016/4.293 &
4.417/4.587 & 4.359 \\
& 1 & 4.382 & 4.526 & 4.146 & 4.493 & 4.375 \\
& 2 & 4.433 & 4.597 & 4.218 & 4.536 & 4.402 \\
$ss\bar{b}\bar{b}$ & 0 & 11.078/11.232 & 11.157/11.254 & 10.777/10.875 &
10.972/11.155 & 10.932 \\
& 1 & 11.099 & 11.199 & 10.820 & 10.986 & 10.939 \\
& 2 & 11.119 & 11.224 & 10.844 & 11.004 & 10.950 \\
$ss\bar{c}\bar{b}$ & 0 & 7.693/7.875 & 7.774/7.960 & 7.394/7.581 &
7.735/7.894 & 7.673 \\
& \multirow{2}{*}{1} & 7.716/7.757 & 7.793/7.872 & 7.414/7.493 &
7.752/7.775 & \multirow{2}{*}{7.683/7.684} \\
& & 7.858 & 7.924 & 7.545 & 7.881 & \\
& 2 & 7.779 & 7.908 & 7.529 & 7.798 & 7.701 \\ \hline\hline
\end{tabular}%
\end{table*}

\medskip \textbf{ACKNOWLEDGMENTS}

W. Z thanks Xiang Liu, Si-Qiang Luo and Hong-Tao An for useful discussions.
D. J thanks Xue-Qian Li and Si-Qiang Luo for useful discussions. This work
is supported by the National Natural Science Foundation of China under Grant
No. 12165017 and No. 12005168.

\medskip

\section*{Appendix A}

\label{AppendixA} \setcounter{equation}{0} \renewcommand{\theequation}{A%
\arabic{equation}} For meson $q_{1}\bar{q}_{2}$(denoted by $M$), baryon $%
(q_{1}q_{2})q_{3}$ (denoted by $B$), and tetraquark systems $q_{1}q_{2}\bar{q%
}_{3}\bar{q}_{4}$(denoted by $T$), the full color wavefunctions, which
respect $SU(3)_{c}$ symmetry, can be written as
\begin{equation}
\phi ^{M}=\frac{1}{\sqrt{3}}\left( r\bar{r}+g\bar{g}+b\bar{b}\right) ,
\label{pcolorM}
\end{equation}%
\begin{equation}
\phi ^{B}=\frac{1}{\sqrt{6}}\left( gbr-bgr+brg-rbg+rgb-grb\right) ,
\label{pcolorB}
\end{equation}%
\begin{equation}
\begin{aligned} \phi_{1}^{T}&=\frac{1}{\sqrt{6}}
\left(rr\bar{r}\bar{r}+gg\bar{g}\bar{g}+bb\bar{b}\bar{b} \right)
+\frac{1}{2\sqrt{6}} \left(rb\bar{b}\bar{r}+br\bar{b}\bar{r} \right.\\
&\left.
+gr\bar{g}\bar{r}+rg\bar{g}\bar{r}+gb\bar{b}\bar{g}+bg\bar{b}\bar{g}+gr%
\bar{r}\bar{g}+rg\bar{r}\bar{g}+gb\bar{g}\bar{b} \right.\\ &\left.
+bg\bar{g}\bar{b}+rb\bar{r}\bar{b}+br\bar{r}\bar{b} \right), \\
\phi_{2}^{T}&=\frac{1}{2\sqrt{3}}
\left(rb\bar{b}\bar{r}-br\bar{b}\bar{r}-gr\bar{g}\bar{r}+rg\bar{g}\bar{r}+gb%
\bar{b}\bar{g}-bg\bar{b}\bar{g} \right.\\ &\left.
+gr\bar{r}\bar{g}-rg\bar{r}\bar{g}-gb\bar{g}\bar{b}+bg\bar{g}\bar{b}-rb%
\bar{r}\bar{b}+br\bar{r}\bar{b}\right), \end{aligned}  \label{pcolorT}
\end{equation}%
respectively. Here, the wavefunction $\phi _{1}^{T}$ in Eq. (\ref{colorT})
corresponds to the configuration $6_{c}\otimes \bar{6}_{c}$ while $\phi
_{2}^{T}$ there corresponds to $3_{c}\otimes \bar{3}_{c}$.

Using the color wavefunctions above and Eq.~(\ref{colorfc}), one can compute
the matrices of color factors. The results can be given explicitly by
\begin{equation}
\left\langle \boldsymbol{\lambda _{1}}\cdot \boldsymbol{\lambda _{2}}%
\right\rangle =-\frac{16}{3},  \label{cfcM}
\end{equation}%
for meson with the wavefunction $(\phi ^{M})$ and
\begin{equation}
\left\langle \boldsymbol{\lambda _{1}}\cdot \boldsymbol{\lambda _{2}}%
\right\rangle =\left\langle \boldsymbol{\lambda _{1}}\cdot \boldsymbol{%
\lambda _{3}}\right\rangle =\left\langle \boldsymbol{\lambda _{2}}\cdot
\boldsymbol{\lambda _{3}}\right\rangle =-\frac{8}{3},  \label{cfcB}
\end{equation}%
for baryon with $(\phi ^{B})$. For tetraquarks with the two-components
wavefunctions $(\phi _{1}^{T},\phi _{2}^{T})$, the matrices of color factors
are
\begin{equation}
\begin{aligned} &\left\langle \boldsymbol{\lambda_{1}}\cdot
\boldsymbol{\lambda_{2}} \right\rangle= \left\langle
\boldsymbol{\lambda_{3}}\cdot \boldsymbol{\lambda_{4}} \right\rangle=
\begin{bmatrix} \frac{4}{3} & 0 \\ 0 & -\frac{8}{3} \end{bmatrix}, \\
&\left\langle \boldsymbol{\lambda_{1}}\cdot \boldsymbol{\lambda_{3}}
\right\rangle= \left\langle \boldsymbol{\lambda_{2}}\cdot
\boldsymbol{\lambda_{4}} \right\rangle= \begin{bmatrix} -\frac{10}{3} &
2\sqrt{2} \\ 2\sqrt{2} & -\frac{4}{3} \end{bmatrix}, \\ &\left\langle
\boldsymbol{\lambda_{1}}\cdot \boldsymbol{\lambda_{4}} \right\rangle=
\left\langle \boldsymbol{\lambda_{2}}\cdot \boldsymbol{\lambda_{3}}
\right\rangle= \begin{bmatrix} -\frac{10}{3} & -2\sqrt{2} \\ -2\sqrt{2} &
-\frac{4}{3} \end{bmatrix}, \end{aligned}  \label{cfcT}
\end{equation}%
all of which are $2\times 2$ matrices in the space of the two-components
wavefunction $(\phi _{1}^{T},\phi _{2}^{T})$.

\medskip

\section*{Appendix B}

\label{AppendixB} \setcounter{equation}{0} \renewcommand{\theequation}{B%
\arabic{equation}} For meson $q_{1}\bar{q}_{2}$(denoted by $M$), baryon $%
(q_{1}q_{2})q_{3}$ (denoted by $B$), and tetraquark systems $q_{1}q_{2}\bar{q%
}_{3}\bar{q}_{4}$(denoted by $T$), one can write the spin wavefunctions for
them, with the help of the Clebsch-Gordan coefficients. The results are
\begin{equation}
\chi _{1}^{M}=\uparrow \uparrow ,\quad \chi _{2}^{M}=\frac{1}{\sqrt{2}}%
\left( \uparrow \downarrow -\downarrow \uparrow \right) ,  \label{pspinM}
\end{equation}%
for the mesons, and
\begin{equation}
\begin{aligned} \chi_{1}^{B}&=\uparrow\uparrow\uparrow, \\
\chi_{2}^{B}&=\sqrt{\frac{2}{3}}\uparrow\uparrow\downarrow-\frac{1}{%
\sqrt{6}}
\left(\uparrow\downarrow\uparrow+\downarrow\uparrow\uparrow\right), \\
\chi_{3}^{B}&=\frac{1}{\sqrt{2}}
\left(\uparrow\downarrow\uparrow-\downarrow\uparrow\uparrow\right),
\end{aligned}  \label{pspinB}
\end{equation}%
for the baryons. For the tetraquark there are six of spin wavefunctions,
\begin{equation}
\begin{aligned} \chi_{1}^{T}&=\uparrow\uparrow\uparrow\uparrow, \\
\chi_{2}^{T}&=\frac{1}{2}
\left(\uparrow\uparrow\uparrow\downarrow+\uparrow\uparrow\downarrow\uparrow-%
\uparrow\downarrow\uparrow\uparrow-\downarrow\uparrow\uparrow\uparrow%
\right), \\ \chi_{3}^{T}&=\frac{1}{\sqrt{3}}
\left(\uparrow\uparrow\downarrow\downarrow+\downarrow\downarrow\uparrow%
\uparrow\right), \\ &-\frac{1}{2\sqrt{3}}
\left(\uparrow\downarrow\uparrow\downarrow+\uparrow\downarrow\downarrow%
\uparrow+\downarrow\uparrow\uparrow\downarrow+\downarrow\uparrow\downarrow%
\uparrow\right) \\ \chi_{4}^{T}&=\frac{1}{\sqrt{2}}
\left(\uparrow\uparrow\uparrow\downarrow-\uparrow\uparrow\downarrow\uparrow%
\right), \\ \chi_{5}^{T}&=\frac{1}{\sqrt{2}}
\left(\uparrow\downarrow\uparrow\uparrow-\downarrow\uparrow\uparrow\uparrow%
\right), \\ \chi_{6}^{T}&=\frac{1}{2}
\left(\uparrow\downarrow\uparrow\downarrow-\uparrow\downarrow\downarrow%
\uparrow-\downarrow\uparrow\uparrow\downarrow+\downarrow\uparrow\downarrow%
\uparrow\right), \end{aligned}  \label{pspinT}
\end{equation}%
which correspond to the states (\ref{spinM}), (\ref{spinB}) and (\ref{spinT}%
), respectively. .

Given the spin wavefunctions above, one can also compute the matrices of
spin factors with the help of Eq.~(\ref{spinfc}). There is one spin matrix
\begin{equation}
\left\langle \boldsymbol{\sigma _{1}}\cdot \boldsymbol{\sigma _{2}}%
\right\rangle =%
\begin{bmatrix}
1 & 0 \\
0 & -3%
\end{bmatrix}%
,  \label{sfcM}
\end{equation}%
for meson in $(\chi _{1}^{M},\chi _{2}^{M})$ space, and three spin matrices
\begin{align}
\left\langle \boldsymbol{\sigma _{1}}\cdot \boldsymbol{\sigma _{2}}%
\right\rangle & =%
\begin{bmatrix}
1 & 0 & 0 \\
0 & 1 & 0 \\
0 & 0 & -3%
\end{bmatrix}%
, \\
\left\langle \boldsymbol{\sigma _{1}}\cdot \boldsymbol{\sigma _{3}}%
\right\rangle & =%
\begin{bmatrix}
1 & 0 & 0 \\
0 & -2 & -\sqrt{3} \\
0 & -\sqrt{3} & 0%
\end{bmatrix}%
, \\
\left\langle \boldsymbol{\sigma _{2}}\cdot \boldsymbol{\sigma _{3}}%
\right\rangle & =%
\begin{bmatrix}
1 & 0 & 0 \\
0 & -2 & \sqrt{3} \\
0 & \sqrt{3} & 0%
\end{bmatrix}%
,
\end{align}%
for baryon in $(\chi _{1}^{B},\chi _{2}^{B},\chi _{3}^{B})$ space. In the
case of tetraquark, there are six spin matrices,
\begin{align}
\left\langle \boldsymbol{\sigma _{1}}\cdot \boldsymbol{\sigma _{2}}%
\right\rangle & =%
\begin{bmatrix}
1 & 0 & 0 & 0 & 0 & 0 \\
0 & 1 & 0 & 0 & 0 & 0 \\
0 & 0 & 1 & 0 & 0 & 0 \\
0 & 0 & 0 & 1 & 0 & 0 \\
0 & 0 & 0 & 0 & -3 & 0 \\
0 & 0 & 0 & 0 & 0 & -3%
\end{bmatrix}%
, \\
\left\langle \boldsymbol{\sigma _{1}}\cdot \boldsymbol{\sigma _{3}}%
\right\rangle & =%
\begin{bmatrix}
1 & 0 & 0 & 0 & 0 & 0 \\
0 & -1 & 0 & \sqrt{2} & -\sqrt{2} & 0 \\
0 & 0 & -2 & 0 & 0 & -\sqrt{3} \\
0 & \sqrt{2} & 0 & 0 & 1 & 0 \\
0 & -\sqrt{2} & 0 & 1 & 0 & 0 \\
0 & 0 & -\sqrt{3} & 0 & 0 & 0%
\end{bmatrix}%
, \\
\left\langle \boldsymbol{\sigma _{1}}\cdot \boldsymbol{\sigma _{4}}%
\right\rangle & =%
\begin{bmatrix}
1 & 0 & 0 & 0 & 0 & 0 \\
0 & -1 & 0 & -\sqrt{2} & -\sqrt{2} & 0 \\
0 & 0 & -2 & 0 & 0 & \sqrt{3} \\
0 & -\sqrt{2} & 0 & 0 & -1 & 0 \\
0 & -\sqrt{2} & 0 & -1 & 0 & 0 \\
0 & 0 & \sqrt{3} & 0 & 0 & 0%
\end{bmatrix}%
, \\
\left\langle \boldsymbol{\sigma _{2}}\cdot \boldsymbol{\sigma _{3}}%
\right\rangle & =%
\begin{bmatrix}
1 & 0 & 0 & 0 & 0 & 0 \\
0 & -1 & 0 & \sqrt{2} & \sqrt{2} & 0 \\
0 & 0 & -2 & 0 & 0 & \sqrt{3} \\
0 & \sqrt{2} & 0 & 0 & -1 & 0 \\
0 & \sqrt{2} & 0 & -1 & 0 & 0 \\
0 & 0 & \sqrt{3} & 0 & 0 & 0%
\end{bmatrix}%
, \\
\left\langle \boldsymbol{\sigma _{2}}\cdot \boldsymbol{\sigma _{4}}%
\right\rangle & =%
\begin{bmatrix}
1 & 0 & 0 & 0 & 0 & 0 \\
0 & -1 & 0 & -\sqrt{2} & \sqrt{2} & 0 \\
0 & 0 & -2 & 0 & 0 & -\sqrt{3} \\
0 & -\sqrt{2} & 0 & 0 & 1 & 0 \\
0 & \sqrt{2} & 0 & 1 & 0 & 0 \\
0 & 0 & -\sqrt{3} & 0 & 0 & 0%
\end{bmatrix}%
, \\
\left\langle \boldsymbol{\sigma _{3}}\cdot \boldsymbol{\sigma _{4}}%
\right\rangle & =%
\begin{bmatrix}
1 & 0 & 0 & 0 & 0 & 0 \\
0 & 1 & 0 & 0 & 0 & 0 \\
0 & 0 & 1 & 0 & 0 & 0 \\
0 & 0 & 0 & -3 & 0 & 0 \\
0 & 0 & 0 & 0 & 1 & 0 \\
0 & 0 & 0 & 0 & 0 & -3%
\end{bmatrix}%
,  \label{sfcT}
\end{align}%
in the subspace of $(\chi _{1}^{T},\chi _{2}^{T},\chi _{3}^{T},\chi
_{4}^{T},\chi _{5}^{T},\chi _{6}^{T})$.

One can then use these factors of color and spin space to find the matrix
representation of the CMI in Eq. (\ref{CMI}) for a given hadronic state. The
color and spin factors are the diagonal elements of the matrices in Eq.~(\ref%
{cfcM}-\ref{cfcT}) and Eq.~(\ref{sfcM}-\ref{sfcT}), respectively. The
off-diagonal elements of these matrices lead to chromomagnetic mixing of
these basis functions given in Appendix A and at the beginning of this
section.

\medskip

\section*{Appendix C}

\label{AppendixC} \setcounter{equation}{0} \renewcommand{\theequation}{C%
\arabic{equation}} Based on Appendices A and B, one can use Eq. (\ref{CMI})
to calculate the matrices of the CMI in the hadronic basis of the
wavefunctions involved in this work. We list some of them whose non-diagonal
elements are nonvanishing. For instance, for the $(\phi ^{B}\chi
_{2}^{B},\phi ^{B}\chi _{3}^{B})$ mixed state of baryons, the CMI matrix is
\begin{equation}
\begin{pmatrix}
\frac{8}{3}C_{12}-\frac{16}{3}C_{13}-\frac{16}{3}C_{23} & -\frac{8\sqrt{3}}{3%
}C_{13}+\frac{8\sqrt{3}}{3}C_{23} \\
-\frac{8\sqrt{3}}{3}C_{13}+\frac{8\sqrt{3}}{3}C_{23} & -8C_{12}%
\end{pmatrix}%
,  \label{mixing23}
\end{equation}
and for the $(\phi _{2}^{T}\chi _{3}^{T},\phi _{1}^{T}\chi _{6}^{T})$ state
of tetraquarks, it is
\begin{equation}
\begin{pmatrix}
\frac{8}{3}(\alpha -\beta ) & 2\sqrt{6}\beta \\
2\sqrt{6}\beta & 4\alpha%
\end{pmatrix}%
,  \label{21mixing36}
\end{equation}

For other cases of tetraquarks, the CMI matrices can be obtained similarly.
They are
\begin{equation}
\begin{pmatrix}
-\frac{8}{3}\theta & -2\sqrt{2}\beta \\
-2\sqrt{2}\beta & -\frac{4}{3}\eta%
\end{pmatrix}%
,  \label{21mixing54}
\end{equation}%
for the tetraquark wavefunction $(\phi _{2}^{T}\chi _{5}^{T},\phi
_{1}^{T}\chi _{4}^{T})$ and
\begin{equation}
\begin{pmatrix}
-\frac{4}{3}(\alpha +5\beta ) & 2\sqrt{6}\beta \\
2\sqrt{6}\beta & -8\alpha%
\end{pmatrix}%
,  \label{12mixing36}
\end{equation}%
for the tetraquark $(\phi _{1}^{T}\chi _{3}^{T},\phi _{2}^{T}\chi _{6}^{T})$%
. In the case of three-dimensional subspace, one can find the CMI matrix to
be
\begin{equation}
\begin{pmatrix}
\frac{4}{3}(2\alpha -\beta ) & \frac{4\sqrt{2}}{3}\delta & 4\delta \\
\frac{4\sqrt{2}}{3}\delta & \frac{8}{3}\eta & -2\sqrt{2}\beta \\
4\delta & -2\sqrt{2}\beta & \frac{4}{3}\beta%
\end{pmatrix}%
,  \label{221mixing245}
\end{equation}%
for the mixed state of $(\phi _{2}^{T}\chi _{2}^{T},\phi _{2}^{T}\chi
_{4}^{T},\phi _{1}^{T}\chi _{5}^{T})$ of tetraquark and
\begin{equation}
\begin{pmatrix}
\frac{4}{3}(2\alpha -\beta ) & -\frac{4\sqrt{2}}{3}\gamma & -4\gamma \\
-\frac{4\sqrt{2}}{3}\gamma & -\frac{8}{3}\theta & -2\sqrt{2}\beta \\
-4\gamma & -2\sqrt{2}\beta & -\frac{4}{3}\eta%
\end{pmatrix}%
,  \label{221mixing254}
\end{equation}%
for the tetraquark state of $(\phi _{2}^{T}\chi _{2}^{T},\phi _{2}^{T}\chi
_{5}^{T},\phi _{1}^{T}\chi _{4}^{T})$, in additional to
\begin{equation}
\begin{pmatrix}
-\frac{2}{3}(2\alpha +5\beta ) & \frac{10\sqrt{2}}{3}\delta & 4\delta \\
\frac{10\sqrt{2}}{3}\delta & -\frac{4}{3}\eta & -2\sqrt{2}\beta \\
4\delta & -2\sqrt{2}\beta & -\frac{8}{3}\theta%
\end{pmatrix}%
,  \label{112mixing245}
\end{equation}%
for the tetraquark state of $(\phi _{1}^{T}\chi _{2}^{T},\phi _{1}^{T}\chi
_{4}^{T},\phi _{2}^{T}\chi _{5}^{T})$. In these matrices, one used $\alpha
=C_{12}+C_{34}$, $\beta =C_{13}+C_{14}+C_{23}+C_{24}$, $\gamma
=C_{13}+C_{14}-C_{23}-C_{24}$, $\delta =C_{13}-C_{14}+C_{23}-C_{24}$, $\eta
=C_{12}-3C_{34}$ and $\theta =3C_{12}-C_{34}$, from Ref.~\cite{Luo:2017eub}.

In the following, we list the expressions for overall binding energy of the
hadrons involved in this work. They are
\begin{equation}
B_{12}+B_{13}+B_{23},  \label{phbinding}
\end{equation}%
for the baryons described by $\phi ^{B}$. For the tetraquarks, the overall
binding energy is
\begin{equation}
-\frac{1}{2}B_{12}+\frac{5}{4}B_{13}+\frac{5}{4}B_{14}+\frac{5}{4}B_{23}+%
\frac{5}{4}B_{24}-\frac{1}{2}B_{34},  \label{ph1binding}
\end{equation}%
for the configuration $\phi _{1}^{T}$, and
\begin{equation}
B_{12}+\frac{1}{2}B_{13}+\frac{1}{2}B_{14}+\frac{1}{2}B_{23}+\frac{1}{2}%
B_{24}+B_{34},  \label{ph2binding}
\end{equation}%
for the configuration $\phi _{2}^{T}$, respectively. Here, the notation $%
B_{ij}$ ($i,j=1,2,3,4$ corresponding to $b$, $c$, $s$ and $n=u, d$) stands for
the binding energies in Eq.~(\ref{Bcs}), and is assumed to be vanish if $%
ij=nn,sn$ or $ss$, in which case there is no short-distance binding in
hadrons \cite{KR:2014gca}.


\begin{thebibliography}{99}

\bibitem{Aaij:2017ueg} LHCb collaboration, R.~Aaij \textit{et al.},
Observation of the doubly charmed baryon $\Xi _{cc}^{++}$, Phys. Rev. Lett.
\textbf{119}, 112001 (2017), arXiv:1707.01621 [hep-ex].

\bibitem{LhcXi:prl18} LHCb collaboration, R. Aaij et al., First observation
of the doubly charmed baryon decay $\Xi _{cc}^{++}\rightarrow $ $\Xi
_{cc}^{+}\pi ^{+}$, Phys. Rev. Lett. 121 (2018) 162002, arXiv:1807.01919
[hep-ex].

\bibitem{LhcXiLf:prl18} LHCb collaboration, R. Aaij et al., Measurement of
the lifetime of the doubly charmed baryon $\Xi _{cc}^{++}$, Phys. Rev. Lett.
121, 052002(2018) arXiv:1806.02744[hep-ex].

\bibitem{LhcXcc:cp20} LHCb collaboration, R. Aaij et al., Measurement of $%
\Xi _{cc}^{++}$ production in pp collisions at $\sqrt{s}$ $=13$ TeV, Chin.
Phys. C44 (2020) 022001, arXiv:1910.11316[hep-ex].

\bibitem{Karliner:2017qjm} M.~Karliner and J.~L.~Rosner, Discovery of
doubly-charmed $\Xi _{cc}$ baryon implies a stable ($bb\bar{u}\bar{d}$)
tetraquark, Phys. Rev. Lett. \textbf{119}, 202001 (2017), arXiv:1707.07666
[hep-ph].

\bibitem{Eichten:2017ffp} E.~J.~Eichten and C.~Quigg,Heavy-quark symmetry
implies stable heavy tetraquark mesons $Q_{i}Q_{j}\bar{q}_{k}\bar{q}_{l}$,
Phys. Rev. Lett. \textbf{119}, 202002 (2017), arXiv:1707.09575 [hep-ph]

\bibitem{Luo:2017eub} S.~Q.~Luo, K.~Chen, X.~Liu, Y.~R.~Liu and S.~L.~Zhu,
Exotic tetraquark states with the $qq\bar{Q}\bar{Q}$ configuration, Eur.
Phys. J. C \textbf{77}, 709 (2017), arXiv:1707.01180 [hep-ph]

\bibitem{LHCb:2020kd} LHC Seminar, $B\to D\bar{D}h$ decays: A new (virtual)
laboratory for exotic particle searches at LHCb, by Daniel Johnson, CERN,
August 11, 2020, \href{https://indico.cern.ch/event/900975/}{%
https://indico.cern.ch/event/900975/}.

\bibitem{Karliner:2020vsi} M.~Karliner and J.~L.~Rosner, First exotic hadron
with open heavy flavor: $cs\bar{u}\bar{d}$ tetraquark, Phys. Rev. D \textbf{%
102}, 094016 (2020), arXiv:2008.05993 [hep-ph]

\bibitem{TccPoly:2021} I. Polyakov, [on behalf of LHCb Collaboration], Talk
at the Euro. Phys. Soc. Conference on High Energy Physics, 29 July
(2021); \href{https://indico.desy.de/event/28202/contributions/105627/attachments/67806/84639/ EPS-HEP\_2021\_Polyakov\_v5.pdf}
{https://indico.desy.de/event/28202/contributions/105627/attachments \\
/67806/84639/ EPS-HEP\_2021\_Polyakov\_v5.pdf}

\bibitem{Fleck:1989mb} S.~Fleck and J.~M.~Richard, Baryons with double
charm, Prog. Theor. Phys. \textbf{82}, 760-774 (1989)

\bibitem{He:2004px} D.~H.~He, K.~Qian, Y.~B.~Ding, X.~Q.~Li and P.~N.~Shen,
Evaluation of spectra of baryons containing two heavy quarks in bag model,
Phys. Rev. D \textbf{70}, 094004 (2004), arXiv:0403301[hep-ph]

\bibitem{Bernotas:2012nz} A.~Bernotas and V.~Simonis, Magnetic moments of
heavy baryons in the bag model reexamined, arXiv:1209.2900 [hep-ph]

\bibitem{KR:2014gca} M.~Karliner and J.~L.~Rosner, Baryons with two heavy
quarks: Masses, production, decays, and detection, Phys. Rev. D \textbf{90},
094007 (2014), arXiv:1408.5877 [hep-ph]

\bibitem{Aliev:2012iv} T.~M.~Aliev, K.~Azizi and M.~Savci,
The masses and residues of doubly heavy spin-3/2 baryons,
J. Phys. G \textbf{40}, 065003 (2013)
doi:10.1088/0954-3899/40/6/065003 [arXiv:1208.1976 [hep-ph]]

\bibitem{Ebert:2004ck} D.~Ebert, R.~N.~Faustov, V.~O.~Galkin and
A.~P.~Martynenko,Semileptonic decays of doubly heavy baryons in the
relativistic quark model,Phys. Rev. D \textbf{70}, 014018 (2004) [erratum:
Phys. Rev. D \textbf{77}, 079903 (2008)], arXiv:hep-ph/0404280 [hep-ph]

\bibitem{Roberts:2007ni} W.~Roberts and M.~Pervin, Heavy baryons in a quark
model, Int. J. Mod. Phys. A \textbf{23}, 2817-2860 (2008), arXiv:0711.2492
[nucl-th]

\bibitem{Albertus:2006ya} C.~Albertus, E.~Hernandez, J.~Nieves and
J.~M.~Verde-Velasco, Static properties and semileptonic decays of doubly
heavy baryons in a nonrelativistic quark model, Eur. Phys. J. A \textbf{32},
183-199 (2007) [erratum: Eur. Phys. J. A \textbf{36}, 119 (2008)],
arXiv:hep-ph/0610030 [hep-ph]

\bibitem{Giannuzzi:2009gh} F.~Giannuzzi, Doubly heavy baryons in a Salpeter
model with AdS/QCD inspired potential, Phys. Rev. D \textbf{79}, 094002
(2009), arXiv:0902.4624 [hep-ph]

\bibitem{Bernotas:2008fv} A.~Bernotas and V.~Simonis, Mixing of heavy
baryons in the bag model calculations, Lith. J. Phys. Tech. Sci. \textbf{48}%
, 127 (2008), arXiv:0801.3570 [hep-ph]

\bibitem{Liu:2018euh} M.~Z.~Liu, Y.~Xiao and L.~S.~Geng, Magnetic moments of
the spin-1/2 doubly charmed baryons in covariant baryon chiral perturbation
theory, Phys. Rev. D \textbf{98}, 014040 (2018), arXiv:1807.00912 [hep-ph]

\bibitem{Gutsche:2011vb} T.~Gutsche, V.~E.~Lyubovitskij, I.~Schmidt and A.~Vega,
Dilaton in a soft-wall holographic approach to mesons and baryons,
Phys. Rev. D \textbf{85}, 076003 (2012), arXiv:1108.0346 [hep-ph].

\bibitem{Gutsche:2017oro} T.~Gutsche, V.~E.~Lyubovitskij and I.~Schmidt,
Tetraquarks in holographic QCD, Phys. Rev. D \textbf{96}, 034030 (2017),
 arXiv:1706.07716 [hep-ph].

\bibitem{Dosch:2016zdv} H.~G.~Dosch, G.~F.~de Teramond and S.~J.~Brodsky,
Supersymmetry Across the Light and Heavy-Light Hadronic Spectrum II,
Phys. Rev. D \textbf{95}, 034016 (2017), arXiv:1612.02370 [hep-ph].

\bibitem{Nielsen:2018ytt} M.~Nielsen, S.~J.~Brodsky, G.~F.~de T\'eramond,
H.~G.~Dosch, F.~S.~Navarra and L.~Zou, Supersymmetry in the Double-Heavy
Hadronic Spectrum, Phys. Rev. D \textbf{98}, 034002 (2018), arXiv:1805.11567 [hep-ph].

\bibitem{Dosch:2020hqm} H.~G.~Dosch, S.~J.~Brodsky, G.~F.~de T\'eramond,
M.~Nielsen and L.~Zou, Exotic states in a holographic theory,
Nucl. Part. Phys. Proc. \textbf{312-317}, 135 (2021), arXiv:2012.02496 [hep-ph].

\bibitem{Faessler:2006ft} A.~Faessler, T.~Gutsche, M.~A.~Ivanov, J.~G.~Korner,
V.~E.~Lyubovitskij, D.~Nicmorus and K.~Pumsa-ard, Magnetic moments of heavy baryons in the relativistic three-quark model, Phys. Rev. D \textbf{73}, 094013 (2006), arXiv:hep-ph/0602193 [hep-ph].

\bibitem{Ali:2017jda} A.~Ali, J.~S.~Lange and S.~Stone, Exotics: Heavy
pentaquarks and tetraquarks, Prog. Part. Nucl. Phys. \textbf{97}, 123
(2017), arXiv:1706.00610 [hep-ph]

\bibitem{Liu:2019zoy} Y.~R.~Liu, H.~X.~Chen, W.~Chen, X.~Liu and S.~L.~Zhu,
Pentaquark and tetraquark states, Prog. Part. Nucl. Phys. \textbf{107},
237-320 (2019), arXiv:1903.11976 [hep-ph]

\bibitem{Moinester:2002uw} M.~A.~Moinester \textit{et al.} [SELEX], First
Observation of Doubly Charmed Baryons, Czech. J. Phys. \textbf{53},
B201-B213 (2003), arXiv:0212029 [hep-ex]

\bibitem{DeGrand:1975cf} T.~A.~DeGrand, R.~L.~Jaffe, K.~Johnson and
J.~E.~Kiskis, Masses and other parameters of the light hadrons, Phys. Rev. D
\textbf{12}, 2060 (1975)

\bibitem{Johnson:1975zp} K.~Johnson, The M.I.T. Bag Model, Acta Phys. Polon.
B \textbf{6}, 865 (1975) MIT-CTP-494.

\bibitem{Strottman:1979qu} D.~Strottman, Multi - Quark Baryons and the MIT
Bag Model, Phys. Rev. D \textbf{20}, 748-767 (1979)

\bibitem{Haxton:1980mc} W.~C.~Haxton and L.~Heller, The heavy
quark-anti-quark potential in the {MIT} Bag model, Phys. Rev. D \textbf{22},
1198 (1980)

\bibitem{Aerts:1980rf} A.~T.~M.~Aerts and L.~Heller, The potential energy of
three heavy quarks in the {MIT} bag model, Phys. Rev. D \textbf{23}, 185
(1981) doi:10.1103/PhysRevD.23.185

\bibitem{Karliner:2017elp} M.~Karliner and J.~L.~Rosner, Quark-level
analogue of nuclear fusion with doubly-heavy baryons, Nature \textbf{551},
89 (2017), arXiv:1708.02547 [hep-ph]

\bibitem{Karliner:2018bms} M.~Karliner and J.~L.~Rosner, Scaling of P-wave
excitation energies in heavy-quark systems, Phys. Rev. D \textbf{98}, 074026
(2018), arXiv:1808.07869 [hep-ph]

\bibitem{Carlson:1982er} C.~E.~Carlson, T.~H.~Hansson and C.~Peterson,
Meson, Baryon and glueball masses in the MIT bag model, Phys. Rev. D \textbf{%
27}, 1556-1564 (1983)

\bibitem{Tanabashi:D18} M.~Tanabashi \textit{et al.} [Particle Data Group],
\textquotedblleft Review of Particle Physics, Phys. Rev. D \textbf{98}, 030001 (2018) and 2021 update

\bibitem{Chodos:1974pn} A.~Chodos, R.~L.~Jaffe, K.~Johnson and C.~B.~Thorn,
Baryon structure in the bag theory, Phys. Rev. D \textbf{10}, 2599 (1974)

\bibitem{Wang:2016dzu} G.~J.~Wang, R.~Chen, L.~Ma, X.~Liu and S.~L.~Zhu,
Magnetic moments of the hidden-charm pentaquark states, Phys. Rev. D \textbf{%
94}, 094018 (2016), arXiv:1605.01337 [hep-ph].

\bibitem{Ebert:2002pp} D.~Ebert, R.~N.~Faustov and V.~O.~Galkin, Properties
of heavy quarkonia and $B_{c}$ mesons in the relativistic quark model, Phys.
Rev. D \textbf{67}, 014027 (2003), arXiv:hep-ph/0210381 [hep-ph]

\bibitem{PACS-CS:2013vie} Y.~Namekawa \textit{et al.} [PACS-CS], Charmed
baryons at the physical point in 2+1 flavor lattice QCD, Phys. Rev. D
\textbf{87}, 094512 (2013), arXiv:1301.4743 [hep-lat]

\bibitem{Brown:2014ena} Z.~S.~Brown, W.~Detmold, S.~Meinel and K.~Orginos,
Charmed bottom baryon spectroscopy from lattice QCD, Phys. Rev. D \textbf{90}%
, 094507 (2014), arXiv:1409.0497 [hep-lat].

\bibitem{Bezginov:2019mdi} N.~Bezginov, T.~Valdez, M.~Horbatsch, A.~Marsman,
A.~C.~Vutha and E.~A.~Hessels, A measurement of the atomic hydrogen Lamb
shift and the proton charge radius, Science \textbf{365}, 1007 (2019)

\bibitem{Xiong:2019umf} W.~Xiong, A.~Gasparian, H.~Gao, \textit{et al.} A
small proton charge radius from an electron--{}proton scattering experiment,
Nature \textbf{575}, 147 (2019)

\bibitem{Aliev:2008ay} T.~M.~Aliev, K.~Azizi and A.~Ozpineci, Magnetic Moments
of Heavy $\Xi_{Q}$ Baryons in Light Cone QCD Sum Rules, Phys. Rev. D \textbf{77},
114006 (2008) doi:10.1103/PhysRevD.77.114006 [arXiv:0803.4420 [hep-ph]].

\bibitem{Aliev:2008sk} T.~M.~Aliev, K.~Azizi and A.~Ozpineci, Mass and Magnetic
Moments of the Heavy Flavored Baryons with J=3/2 in Light Cone QCD Sum Rules,
Nucl. Phys. B \textbf{808}, 137-154 (2009) doi:10.1016/j.nuclphysb.2008.09.018
[arXiv:0807.3481 [hep-ph]].

\bibitem{Azizi:2021aib} K.~Azizi and U.~\"Ozdem, Magnetic dipole moments of
the $T_{cc}^+$ and $Z_V^{++}$ tetraquark states, [arXiv:2109.02390 [hep-ph]].

\bibitem{Lu:2020rog} Q.~F.~L\"{u}, D.~Y.~Chen and Y.~B.~Dong, Masses of
doubly heavy tetraquarks $T_{QQ^{\prime }}$ in a relativized quark model,
Phys. Rev. D \textbf{102}, 034012 (2020), arXiv:2006.08087 [hep-ph]

\bibitem{Ebert:2007rn} D.~Ebert, R.~N.~Faustov, V.~O.~Galkin and W.~Lucha,
Masses of tetraquarks with two heavy quarks in the relativistic quark
model,Phys. Rev. D \textbf{76}, 114015 (2007), arXiv:0706.3853 [hep-ph]



\end{thebibliography}
\end{document}